\documentclass[amsmath, amssymb, twocolumn, aps, prx, superscriptaddress]{revtex4-1}

\usepackage{mathtools}
\usepackage{mathrsfs, amsmath, wasysym}
\usepackage{bbold}
\usepackage[mathscr]{eucal}
\usepackage{xfrac}
\usepackage{graphicx}
\usepackage{dcolumn}
\usepackage{bm}
\usepackage{color}
\usepackage{tabularx}
\usepackage{natbib}
\usepackage[colorlinks,linkcolor=blue,citecolor=blue,urlcolor=blue]{hyperref}
\usepackage[normalem]{ulem}
\usepackage[all]{hypcap}
\usepackage[dvipsnames,usenames,table]{xcolor}
\usepackage{braket,mleftright}

\newcommand{\nocontentsline}[3]{}
\newcommand{\tocless}[2]{\bgroup\let\addcontentsline=\nocontentsline#1{#2}\egroup}

\newcommand{\bk}{{\bf k}}

\newcommand{\bq}{{\bf q}}

\newcommand{\bz}{{\bf z}}

\newcommand{\bA}{{\bf A}}

\newcommand{\bb}{{\bf b}}
\newcommand{\bB}{{\bf B}}
\newcommand{\bC}{{\bf C}}
\newcommand{\bE}{{\bf E}}

\newcommand{\bD}{{\bf D}}

\newcommand{\bP}{{\bf P}}

\newcommand{\br}{{\bf r}}

\newcommand{\bS}{{\bf S}}

\newcommand{\bv}{{\bf v}}

\newcommand{\bw}{{\bf w}}

\newcommand{\bn}{{\bf n}}

\newcommand{\bM}{{\bf M}}

\newcommand{\bh}{{\bf h}}

\newcommand{\btau}{\boldsymbol{\tau} }
\newcommand{\bsigma}{\boldsymbol{\sigma} }

\newcommand{\bLambda}{\boldsymbol{\Lambda} }
\newcommand{\bGamma}{\boldsymbol{\Gamma} }

\newcommand{\be}{\begin{equation}}
\newcommand{\ee}{\end{equation}}
\newcommand{\beg}{\begin{gather}}
\newcommand{\eeg}{\end{gather}}
\newcommand{\beq}{\begin{eqnarray}}
\newcommand{\eeq}{\end{eqnarray}}
\newcommand{\bea}{\begin{align}}
\newcommand{\eea}{\end{align}}
\newcommand{\beqq}{\begin{eqnarray*}}
\newcommand{\eeqq}{\end{eqnarray*}}

\newcommand{\up}{\uparrow}
\newcommand{\down}{\downarrow}

\newcommand{\ve}{{\varepsilon}}

\begin{document}

\title{Berry phase polarization and orbital magnetization responses of insulators: \\
Formulas for generalized polarizabilities and their application}

\author{J. W. F. Venderbos}
\affiliation{Department of Physics, Drexel University, Philadelphia, PA 19104, USA}%
\affiliation{Department of Materials Science \& Engineering, Drexel University, Philadelphia, PA 19104, USA \looseness=-1}%

\begin{abstract}
Condensed matter physics is often concerned with determining the response of a solid to an external stimulus. This paper revisits and extends the microscopic formalism for calculating response coefficients---here referred to as (generalized) polarizabilities---in crystalline electronic insulators. The main focus is on the Berry phase polarization and orbital magnetization, for which we obtain general formulas describing the linear response to an arbitrary (but static and uniform) perturbation. The response of an arbitrary lattice-periodic observable (e.g. spin, layer pseudospin) to electric and magnetic fields is also examined, and serves as a basis for mircoscopically establishing Maxwell relations between conjugate generalized polarizabilities. We furthermore introduce and examine the notion of Berry curvature or Hall vector polarizability, i.e., the response of the Berry curvature to a general perturbation, and show how it relates to Berry phase polarization and orbital magnetization responses. For all polarizabilities considered, we obtain simplified formulas applicable to two- and four-band models, expressed directly in terms of the Hamiltonian and the perturbation. Three specific applications of these formulas are discussed: (i) a computation of the magnetoelectric polarizabilities of model antiferromagnets in one and two dimensions; (ii) a general proof of (quasi)topological signatures in the polarizabilities of Dirac fermions in two dimensions; (iii) a calculation of the strain-induced Berry curvature polarizability in an altermagnet. 
\end{abstract}

\date{\today}

\maketitle

\section{Introduction \label{sec:intro}}

A primary objective of condensed matter physics is to understand and predict how a solid responds to the application of external fields. Classic examples of the latter are electric and magnetic fields, but the study of response properties extends to more general external perturbations such as applied strain or temperature gradients. From a macroscopic perspective, a basic understanding of the response properties of solids can be obtained from the appropriate constitutive equations. Textbook examples include the linear polarization response of an insulator to an applied electric field, or the magnetization response to a magnetic field, but more interesting (and less common) examples are response equations which relate quantities of distinct symmetry, such as magnetoelectric or piezomagnetic response equations~\cite{Fiebig:2005pR123,Spaldin:2005p391,Moriya:1959p73,Borovik:1960p786,Dzialoshinskii:1958p621}. Responses of the latter type highlight a general and important principle: the response coefficients defining a response equation encode key information on the symmetry properties of a solid state system. Determining the symmetry-allowed linear or nonlinear response coefficients is therefore a powerful approach to establishing the nature of an ordered state. Identifying a specific pattern of broken and preserved symmetries through an analysis of the allowed responses is often a key step towards understanding unconventional ordering phenomena~\cite{Hayami:2018p165110,Yatsushiro:2021p054412,Yatsushiro:2021p054412}. The spontaneous breaking of inversion symmetry, for instance, may be inferred from temperature dependent nonlinear optical responses~\cite{Fiebig:1994p2127,Fiebig:2005p96,Zhao:2016p32,Zhao:2016p250,Harter:2017p295,Chu:2020p027601,Ni:2021p782}. In a similar way, magnetoelectric responses are indicative of a particular type of magnetic ordering~\cite{Fiebig:2005pR123,Spaldin:2005p391,Khomksii:2006p1,Eerenstein:2006p759,Cheong:2007p13,Tokura:2014p076501,Barone:2015p143}.
 
Further insight into the response properties of solids can be gained by calculating the symmetry-allowed response functions in terms of energies and wave functions obtained from an appropriate microscopic model. This can provide a deeper and more microscopic  understanding of the---intrinsic or extrinsic---physical mechanisms underlying the expected response, and also serves as a basis for quantitative comparison with experimental observations. 

The purpose of this paper is to revisit the microscopic calculation of polarization and magnetization responses in insulating periodic systems~\cite{Resta:1992p51,King-Smith:1992p1651,Vanderbilt:1993p4442,Resta:1994p899,Thonhauser:2005p137205,Ceresoli:2006p024408}, and to develop a general perturbative framework for computing the associated polarizabilities. The notion of polarizability is used here in a general and broad sense: it refers to the quantity describing the linear response of an electronic insulator to an arbitrary lattice-periodic perturbation. The goal is to obtain a formalism that may be readily applied to many specific settings, and to expose relationships between different types of polarizabilities. Specific examples that will be discussed are magnetoelectric polarizabilities, as well as responses to external strain. 

The starting point of our approach is the microscopic expression for the electric polarization or the magnetization of a band insulator, which is given in terms of band energies and eigenstates of a Bloch Hamiltonian. These expressions are then expanded in a small perturbative parameter representing the external field, such that the response to first order in the perturbation defines the linear (electric or magnetic) polarizability. The core underlying assumption is that a polarization or magnetization is symmetry forbidden in the unperturbed ground state, but is allowed once the external field (i.e., the perturbation) is applied. 

Our main focus is on the Berry phase polarization and the orbital magnetization response. The microscopic recipe for computing these is provided by the ``modern theory'' of polarization and orbital magnetization~\cite{Resta:1994p899,Thonhauser:2005p137205,Xiao:2005p137204,Ceresoli:2006p024408,Xiao:2010p1959}, and conceptually relies on the Wannier state construction of energy bands~\cite{Marzari:2012p1419}. The expressions for the polarization and the orbital magnetization are provided in terms of band energies and (derivatives of) Bloch eigenstates, and reflect band geometric properties encoded in the Bloch wave functions. A key insight from the modern theory of polarization is that only changes of polarization are gauge-invariant and well-defined, and that the derivative of the polarization with respect to an adiabatic parameter is given by a band geometric curvature~\cite{Resta:1992p51,King-Smith:1992p1651}. This understanding of polarization differentials forms the basis for the microscopic description of electric polarizabilities.

In the case of the orbital magnetization, the question of external field responses was previously addressed in the context of magnetoelectric properties of band insulators~\cite{Malashevich:2010p053032}, motivated in large part by the predicted topological magnetoelectric effect in topological insulators~\cite{Qi:2008p195424,Essin:2009p146805}. The object of interest was a microscopic theory of the orbital magnetoelectric polarizability (OMP), defined as the orbital magnetization response to an electric field~\cite{Malashevich:2010p053032} (or, equivalently, the polarization response to a magnetic field~\cite{Essin:2010p205104}). This work builds on these and other previous efforts~\cite{Mahon:2019p235140,Mahon:2020p033126} and generalizes the notion of orbital magnetic polarizability by deriving expressions for the orbital magnetization response to arbitrary perturbations. 

General thermodynamic arguments suggest that the polarization and magnetization responses described by the modern theory are related via Maxwell relations to dual responses produced by electric or magnetic fields. 
For instance, the polarization response to a Zeeman magnetic field (i.e., the polarization response originating from the Zeeman coupling) is expected to equal the spin magnetization response to an electric field~\cite{Garate:2009p134403,Gao:2018p134423,Manchon:2019p035004,Shitade:2018p020407,Shitade:2019p024404,Xiao:2022p086602}. In order to demonstrate that such Maxwell relations hold microscopically (Sec.~\ref{sec:apply-ME}), we also examine pseudospin polarization and spin magnetization responses. These refer to responses that manifest as ground state expectation values of an internal electric or magnetic dipole degree of freedom. The case of spin is well-known, but internal electric pseudospin degrees are less common. A notable example is the layer degree of freedom in two-dimensional (2D) multilayer materials, which couples to an electric displacement field applied perpendicular to the 2D layers~\cite{Pesin:2012p409,Xu:2014p343,Gao:2020p077401,Zheng:2024p8017,Fan:2024p7997,Hu:2024pL201403,Hu:2025arXiv}. The significance of the layer degree of freedom for a variety of phenomena observed in multilayer systems is steadily gaining appreciation~\cite{Balents:2020p725,Andrei:2020p1265,Andrei:2021p201,Mak:2022p686,Pantaleon:2023p304,Bernevig:2024p38,Nuckolls:2024p460,Li:arXiv2025,Burch:2018p47,Gibertini:2019p408,Xie:2022p9525}, and is likely to increase further as the 2D materials frontier expands.  

In addition to electric and magnetic polarizabilities with a familiar physical interpretation (i.e., polarization and magnetization responses), we also introduce and examine a polarizability of a different kind: the Berry curvature---or Hall vector---polarizability. The Hall vector polarizability describes how the Berry curvature of a band or set of bands responds to a perturbation, and is therefore relevant for responses and properties which rely on the Berry curvature distribution~\cite{Xiao:2010p1959,Nagaosa:2010p1539}. Here we show how the Hall vector polarizability relates to the Berry phase polarization and orbital magnetization responses described by the modern theory. We will demonstrate in particular that the Hall vector polarizability can be recast as a generalized curvature dipole, i.e., a generalization of the Berry curvature dipole, and that under certain conditions the interband term of orbital magnetic polarizability can be rewritten in terms of the Hall vector polarizability. 

Our derivation of the polarizability expressions relies on a projector-based description of energy bands eigenstates, i.e., a description in terms of projectors onto the occupied and unoccupied bands~\cite{Graf:2021p085114,Mitscherling:2025p085104}. This has a number of advantages. The first is that such a description makes the gauge invariance of the obtained expressions manifest. The second and more practically useful advantage is that in some cases the form of the projectors follows in a straightforward manner from the Hamiltonian, which has the appealing benefit that the polarizability expressions can be readily recast in terms of the parametrization of the Hamiltonian. This is the case for two-band models, for instance, and also for a special class of four-band models. Since such models are often used as effective models for a set of relevant low-energy bands, or as minimal models for the essential electronic structure of materials, we determine the simplified polarizability formulas for two- and four-band models. For general $N$-band models it is still possible to obtain formulas in terms parametrizations of the projectors, but the recipe for relating these to the parametrization of the Hamiltonian becomes more involved as $N$ increases~\cite{Graf:2021p085114}.

In the second part of this paper we then consider three specific applications of these general formulas. In Secs.~\ref{sec:apply-1D} and \ref{sec:2D-bilayer-AFM} we study minimal models of insulating magnetoelectric antiferromagnets in one~\cite{Yanase:2014p014703} and two dimensions~\cite{Venderbos:arXiv2025,Radhakrishnan:arXiv2025} (see Refs.~\onlinecite{Burch:2018p47} and \onlinecite{Gibertini:2019p408} for reviews of magnetism in reduced dimensions), respectively, and show how the obtained formulas readily yield the magnetoelectric polarizabilities~\cite{Yanase:2014p014703,Hayami:2015p064717,Hayami:2022p123701,Yatsushiro:2022p155157}. In Sec.~\ref{sec:2D-Dirac} we turn to a general study of the response properties of electronic systems in two dimensions that can be described by a low-energy continuum Dirac model~\cite{Vafek:2014p83}. Within the general setting of such a model we show that the Berry phase polarization and orbital magnetization responses carry signatures of nontrivial topology, as evidenced by the (quasi)topological nature of the corresponding polarizabilities~\cite{Ramamurthy:2015p085105}. We discuss how this result connects to and extends previous work on topological response theory~\cite{Qi:2008p195424,Hughes:2011p245132,Ramamurthy:2015p085105}. Finally, in Sec.~\ref{sec:apply-altermagnet} we examine the notion of Hall vector polarizability in the context of a specific microscopic model for an altermagnet in two dimensions~\cite{Smejkal:2022p031042,Smejkal:2022p040501,Brekke:2023p224421,Antonenko:2025p096703}. Altermagnets are known to be premier venues for strain engineering~\cite{Ma:2021p2846,Aoyama:2024pL041402,Takahashi:2025p184408,Belashchenko:2025p086701,Naka:2025p083702,Khodas:arXiv2025,Zhai:2025p174411,Fu:arXiv2025} and here we show in detail how the Hall vector polarizability captures the way in which the Berry curvature responds to strain~\cite{Takahashi:2025p184408,Smolenski:arXiv2025}.


\section{General approach to calculating polarizabilities}
\label{sec:problem}

We begin by describing the general perturbative approach taken in this work. This will provide the basis for the derivations that follow and will introduce the requisite notation. 

The conceptual starting point is a Bloch Hamiltonian of the form
\be
\mathcal H = H_0 + \lambda W, \label{eq:H-Bloch}
\ee
where $H_0 = H_0(\bk)$ is the Bloch Hamiltonian of an unperturbed periodic crystal and $W = W(\bk)$ describes a perturbation corresponding to an external field $\lambda$. We will formally treat $\lambda$ as a dimensionless field, such that $W$ has dimension energy, but will trivially reinstate the appropriate dimension in any practical setting or application. For the purpose of notational simplicity any dependence on the crystal momentum $\bk$ is suppressed in Eq.~\eqref{eq:H-Bloch}; we will generally adhere to this convention throughout this work. We take $\mathcal H$ to describe an $N$-band system with energy bands denoted $\mathcal E_{\bk n}$ and Bloch eigenstates denoted $|u_{\bk n}\rangle $. Here $n$ is the band index. We denote the band energies and eigenstates corresponding to the unperturbed Hamiltonian $H_0$ as $\varepsilon_{\bk n}$ and $|u^0_{\bk n}\rangle $. Since we will generally suppress the momentum dependence in what follows, we will simply write $\mathcal E_{ n}$ and $|u_{ n}\rangle $, as well as $\varepsilon_{ n}$ and $|u^0_{ n}\rangle $. 

Since our focus is on the response properties of insulators, we assume that the spectrum of $H_0$ has an energy gap and that its ground state is defined by $N_{\text{occ}}$ filled bands. We further assume that the symmetries of $H_0$ forbid both a polarization $\bP$ and a uniform magnetization $\bM$ in the ground state. The perturbation $W$ activated by the external field represents a symmetry breaking term and the motivating assumption is that $W$ breaks the symmetries of $H_0$ in such a way that either a polarization or magnetization (or perhaps both) are allowed. To determine the polarization or magnetization response to $\lambda$ we set out to compute the polarizabilities $\partial \bP/\partial \lambda $ and $\partial \bM/\partial \lambda $ (evaluated at $\lambda =0 $) by applying standard perturbation theory. 

The derivation of the polarizabilities is based on non-degenerate perturbation theory. The assumption of a non-degenerate unperturbed energy spectrum is not essential for obtaining the final results, but it offers the most straightforward route to establishing these results. Furthermore, as will become evident, the final expressions for the relevant polarizabilities only involve energy differences of occupied and unoccupied bands, and the occupied and unoccupied subspaces are manifestly separated by an energy gap. Therefore, degeneracies within the occupied or unoccupied subspaces are not a cause for concern when evaluating the polarizability formulas. Nonetheless, all derivations can be appropriately adapted to the framework of degenerate perturbation theory in case of band degeneracies. 

A key object of interest in determining the polarizabilities are the corrections to the unperturbed eigenstates. Within standard perturbation theory the eigenstate $ |  u_n\rangle $ is expanded in terms of the unperturbed eigenstates as
\be
 |  u_n\rangle =  |  u_n^0\rangle + \lambda \sum_{l\neq n} \frac{\langle  u_l^0|W  | u_n^0 \rangle}{\varepsilon_{n}-\varepsilon_l}  |  u_l^0\rangle + \mathcal O(\lambda^2), \label{eq:u_n}
\ee
where $|  u_n^0\rangle$ denotes the $n$-th  unperturbed eigenstate corresponding to the unperturbed band energies $\varepsilon_{n}$. Rather than work with corrections to eigenstates, we will find it useful to instead work with corrections to the projectors onto the corresponding eigenspaces. We define $\mathcal P_n =  |  u_n\rangle  \langle u_n |$ as the projector onto $ |  u_n\rangle $, and this projector is formally expanded as 
\be
\mathcal P_n = \mathcal P^{(0)}_n+ \lambda \mathcal P^{(1)}_n + \mathcal O(\lambda^2). \label{eq:P_n}
\ee
Here $\mathcal P^{(0)}_n = |  u^0_n\rangle  \langle u^0_n |$ is the projector onto the unperturbed state $ |  u^0_n\rangle$, and in what follows we will denote the unperturbed projectors as $ P_n$. Hence $\mathcal P^{(0)}_n = |  u^0_n\rangle  \langle u^0_n | \equiv P_n$. The first order correction to the projector is given by 
\be
\mathcal P^{(1)}_n = \sum_{l\neq n} \frac{P_n W P_l+P_l W P_n}{\varepsilon_{n}-\varepsilon_{l}}. \label{eq:P^(1)_n}
\ee
Throughout this paper we will label the occupied bands by $n, n',n''$, etc., and the unoccupied bands by $m,m',m''$, etc. As a result, $\sum_n$ and $\sum_m$ will denote sums over occupied and unoccupied bands, respectively. For instance, $\mathcal P = \sum_n \mathcal P_n$ is defined as the projector onto the full occupied subspace, and $\mathcal Q = \sum_m \mathcal P_m = \mathbb{1} - \mathcal P$ is the projector onto the unoccupied subspace. Correspondingly, $P = \sum_n P_n$ and $ Q = \sum_m P_m = \mathbb{1} -   P$ are the projectors onto the occupied and unoccupied subspaces of the unperturbed Hamiltonian $H_0$. Given this notation convention, the first order correction to the projector onto the subspace of occupied bands is given by
\be
\mathcal P^{(1)} = \sum_{ n,m } \frac{\{ P_n ,\{ W, P_m \}\}}{\varepsilon_{n}-\varepsilon_{m}}. \label{eq:P^(1)}
\ee
Note that it is straightforward to show that $\mathcal Q^{(1)} =-\mathcal P^{(1)} $.

In this paper we will make frequent use of a convenient and elegant description of $N$-band systems proposed in Ref.~\onlinecite{Graf:2021p085114}. Rather than working with the eigenstates, this approach is focused on determining the eigenstate projectors of an $N$-band Hamiltonian. Within such a projector-based formulation of $N$-band systems, the unperturbed Hamiltonian is written as $H_0 =h_0\mathbb{1}+ \bh \cdot \bLambda$, where $\bh$ is a real $(N^2-1)$-component vector and $\bLambda$ is a vector collecting the $N^2-1$ generators of the SU($N$) Lie group (i.e., the basis of the corresponding Lie algebra). Note that $h_0 = h_{0,\bk}$ and $\bh = \bh_\bk$ are functions of momentum, but, as mentioned above, we generally suppress the momentum dependence. The generators satisfy the commutation and anti-commutation relations
\begin{align}
[\Lambda_\alpha  ,\Lambda_\beta] & = 2i f_{\alpha\beta\gamma} \Lambda_\gamma, \label{eq:L-commute} \\
 \{ \Lambda_\alpha  ,\Lambda_\beta \} &= \frac{4}{N}\delta_{\alpha\beta} \mathbb{1}+ 2d_{\alpha\beta\gamma} \Lambda_\gamma  ,\label{eq:L-anticommute}
\end{align}
where $f_{\alpha\beta\gamma} $ and $d_{\alpha\beta\gamma} $ are the structure constants. The projector onto the eigenspace of the $n$-th band then takes the form
\be
P_n = \frac{1}{N}\mathbb{1} + \frac12 \bb_n \cdot \bLambda \label{eq:SU(N)-Pn},
\ee
where $\bb_n$ may be determined from the energies $\varepsilon_{n}$ and $\bh$, using the recipe provided in Ref.~\onlinecite{Graf:2021p085114}. (Note that this discussion assumes the absence of any manifest global band degeneracies.) 

Any operator can be expanded in the generators $\bLambda$ and this is in particular true for the perturbation $W$. Writing $W = w_0\mathbb{1}+ \bw \cdot \bLambda$, where $\bw$ is a real $(N^2-1)$-component vector parametrizing the perturbation, we will use Eq.~\eqref{eq:SU(N)-Pn} to obtain expressions for the polarizabilities of interest in terms of SU($N$) vectors such as $\bh$, $\bb_n$, and $\bw$. (In cases where only interband matrix elements are required, neither $h_0$ nor $w_0$ enter the expressions for the polarizabilities.)

Special attention will be given to the cases $N=2$ and $N=4$. The case of two-band models ($N=2$) is particularly simple, since the generators reduce to the three Pauli matrices $\bsigma = (\sigma^x,\sigma^y,\sigma^z)$. A generic two-band Hamiltonian can then be written as $H_0 =h_0\mathbb{1}+ \bh \cdot \bsigma$, where $\bh = (h^x,h^y,h^z)$ is a three-component vector, and the projector onto the occupied valence band takes the form $P = \frac12(\mathbb{1}  - \bh \cdot \bsigma/|\bh|)$. Note that the assumption of an insulating energy gap means that $|\bh|>0$ for all $\bk$. 

In the case of $N=4$, we will focus on a special class of four-band models which satisfy an additional symmetry constraint. Specifically, we will consider the most general four-band Hamiltonian with $\mathcal I \mathcal T$ symmetry, where $\mathcal I$ is inversion and $\mathcal T$ is time-reversal. The product symmetry $\mathcal I \mathcal T$ mandates that all energy bands are manifestly twofold degenerate (Kramers degeneracy), such that the most general Hamiltonian can always be written as $H_0 = \bh \cdot \bGamma$ (up to a term proportional to the identity). Here $\bGamma$ is a vector which collects five mutually anticommuting (and Hermitian) $\Gamma$-matrices and $\bh$ is therefore a real five-component vector~\cite{Murakami:2004p235206}. This class of four-band models is of particular practical relevance and its description in terms of anticommuting $\Gamma$-matrices furthermore has appealing formal properties which can be exploited in determining responses. For more details we refer to Appendix~\ref{app:4-band}.

\section{Electric polarizability \label{sec:electric}}

In this section we first consider electrical polarizabilities. By this we mean the electric polarization response to the perturbation $W$ introduced in Eq.~\eqref{eq:H-Bloch}. We discuss two types of polarization responses: the Berry phase polarization and the so-called ``pseudospin'' polarization. The former is the fruit of the modern theory of polarization, which in the introduction we somewhat colloquially referred to as ``orbital'' polarization, and the latter refers to the ground state expectation value of an internal electric dipole moment. An example is the layer dipole moment in 2D bilayer or multilayer materials. 

\subsection{Berry phase polarization \label{ssec:Berry-P}}

The modern theory of polarization states that the differential change in polarization $\bP$ in response to an infinitesimal change of $\lambda$ is given by \cite{Resta:1994p899} \footnote{The use of the symbol $P$ for both polarization and projectors is somewhat unfortunate. However, departure from firmly established convention in either case would be more unfortunate. Context and subscripts should leave the meaning unambiguous.}
\be
 \partial_\lambda P_a = e  \int[d\bk] \,  \Omega_{a\lambda}, \label{eq:dP}
\ee
where $a=x,y,z$ labels the Cartesian coordinates and $\partial_\lambda  = \partial/\partial\lambda$. Here we have further defined $[d\bk] \equiv d^D \bk / (2\pi)^D$ as the integral measure in $D$ dimensions, and the integration is over the Brillouin zone. The integrand $ \Omega_{a\lambda}$ is the Berry curvature and is given by 
\be
\Omega_{a\lambda} =  -2 \text{Im}\text{Tr}[  \mathcal P \partial_a  \mathcal P\partial_\lambda \mathcal P ] = -\sum_n 2 \text{Im}   \langle  \partial_a u_n | \partial_\lambda u_n\rangle ,  \label{eq:Omega_al}
\ee
where $\partial_a  = \partial/\partial k_a$ and $\mathcal P$ is the projector onto the subspace of occupied bands, as introduced in the previous section. (Recall that the momentum dependence is generally suppressed.) For our purpose the expression of the curvature in terms of projectors is most useful. To obtain the linear polarizability, we simply make the substitution $\partial_\lambda \mathcal P \rightarrow \mathcal P^{(1)}$ in \eqref{eq:Omega_al}, with $\mathcal P^{(1)}$ given by Eq.~\eqref{eq:P^(1)}, and find 
\be
\partial_\lambda P_a  =  e \int [d\bk]\, \sum_{n,m}  \frac{2 \text{Im}\text{Tr}[P_n(\partial_a P_{m})   W  ]}{\varepsilon_{n}-\varepsilon_m} ,  \label{eq:dP-2}
\ee
where we recall that the sum on $n$ ($m$) is over the occupied (unoccupied) bands (see Sec.~\ref{sec:problem}). This expression may be recast in terms of velocity matrix elements as 
\be
 \partial_\lambda P_a  =  -e  \int [d\bk] \sum_{n,m}  \frac{2\text{Im}[ v_a^{nm} W^{mn}]}{(\varepsilon_{n}-\varepsilon_m)^2}. \label{eq:dP-3}
\ee
where $v_a^{nm} =\langle  u^0_n |\partial_a H_0  | u^0_m \rangle $ are the (unperturbed) matrix elements of the velocity operator $v_a = \partial_a H_0$, and $ W^{mn}=\langle  u^0_m|W  | u^0_n \rangle$. 

The general expressions \eqref{eq:dP-2} and \eqref{eq:dP-3} represent the first main results of this section. We now determine the form these general expressions reduce to when applied to the $N$-band model description discussed in Sec.~\ref{sec:problem}. We begin by considering $N=2$ and $N=4$, and then turn to the general case. In Appendix \ref{app:berry-connection} we further discuss how the Berry connection polarizability follows from Eqs.~\eqref{eq:dP-2} and \eqref{eq:dP-3} when the perturbation is an electric field~\cite{Gao:2014p166601,Xiao:2022p086602}. 

{\it Application to two-band models.} First, consider a general two-band Hamiltonian of the form $H_0 =h_0 \mathbb{1}+ \bh \cdot \bsigma $, where $\bsigma = (\sigma^x,\sigma^y,\sigma^z)$ are a set of Pauli matrices (see Sec.~\ref{sec:problem}). The energy spectrum has two branches $ \ve_{1,2} =h_0 \mp  |\bh|$, which are separated by an energy gap (by assumption). We assume the lower band $\ve_{1} $ is occupied and the projector onto this band is given by $P = \frac12(\mathbb{1}  - \bh \cdot \bsigma/|\bh|)$. The perturbation $W$ can be parametrized as $W =w_0 \mathbb{1} + \bw \cdot \bsigma$. Substituting these expressions into Eq.~\eqref{eq:dP-2} yields
\be
\partial_\lambda P_a  =  e \int [d\bk]    \frac{\bw \cdot \bh \times \partial_a \bh }{2|\bh|^3 }  \label{eq:dPdL-2-band},
\ee
from which it follows that the polarizability can be expressed directly in terms of the Hamiltonian (through $\bh$), the perturbation (through $\bw$), and the energy $|\bh|$. Note that the polarizability does not depend on $h_0$ and $w_0$. In Sec.~\ref{sec:apply-1D} we discuss examples of using this formula in the context of a specific microscopic model. 
 
{\it Application to four-band models.} Next, we consider the special class of four-band models introduced in Sec.~\ref{sec:problem}, i.e., four-band models with the additional requirement of $\mathcal I \mathcal T$ symmetry. The product of $\mathcal I $ and $\mathcal T$ mandates a manifest twofold degeneracy of all energy bands, such that the most general Hamiltonian compatible with $\mathcal I \mathcal T$ can always be written as $H_0 = \bh \cdot \bGamma$ (up to a term proportional to the identity, which is unimportant in this instance). Here $\bh$ is a real five-component vector and $\bGamma$ is a vector which collects five mutually anticommuting $\Gamma$-matrices which satisfy $\{\Gamma_\alpha,\Gamma_\beta \} = 2\delta_{\alpha\beta}\mathbb{1}$. As a result of this form of $H_0$, the energy bands and the associated eigenspace projectors are given by expressions similar to those of the two-band model. The only difference is that $\bh$ is now a five-component vector. In particular, the spectrum has two branches $ \ve_{1,2} = \mp  |\bh|$, and the projector onto the---now twofold degenerate---occupied band is $P = \frac12(\mathbb{1}  - \bh \cdot \bGamma/|\bh|)$. 

To compute the polarizability using Eq.~\eqref{eq:dP-2}, we require an expression for the perturbation $W$ in terms of the $\Gamma$-matrices. The most general form of the perturbation is $W = w^\alpha\Gamma_\alpha + w^{\alpha\beta}\Sigma_{\alpha\beta}$ (ignoring a uniform term $w_0 \mathbb{1}$), where $\Sigma_{\alpha\beta} = [\Gamma_\alpha,\Gamma_\beta]/2i= - \Sigma_{\beta\alpha}$~\cite{Murakami:2004p235206} (see Appendix \ref{app:4-band} for details). Here $w^\alpha$ is a real five-component vector, like $\bh$, and $w^{\alpha\beta}$ is a real ten-component antisymmetric tensor ($w^{\alpha\beta}=-w^{\beta\alpha}$).  Substituting all this into \eqref{eq:dP-2}, we obtain
\be
\partial_\lambda P_a  =   e \int [d\bk]    \frac{ 2 h^\alpha( \partial_a h^\beta) w^{\alpha\beta} }{|\bh|^3 }  \label{eq:dPdL-4-band},
\ee
which is the generalization of Eq.~\eqref{eq:dPdL-2-band} to four-band models with $\mathcal I \mathcal T$ symmetry. We note that the polarizability is fully determined by $\bh$ and the piece of the perturbation parametrized by $w^{\alpha\beta}$. 

{\it General $N$-band models.} Finally, we proceed to applying Eq.~\eqref{eq:dP-2} to the general $N$-band model introduced at the end of Sec.~\ref{sec:problem}. It should be noted that this formulation of $N$-band models assumes there are no manifest band degeneracies, such that each band can be uniquely represented by a vector $\bb_n$, as defined in Eq.~\eqref{eq:SU(N)-Pn}. With that assumption and given the form of $P_{n}$ and $P_m$, it is straightforward to show that the polarizability of Eq.~\eqref{eq:dP-2} reduces to
\be
\partial_\lambda P_a  =  e \int [d\bk]\, \sum_{n,m}  \frac{ \partial_a \bb_n \times \bb_m \cdot  \bw   }{\varepsilon_{n}-\varepsilon_m} ,   \label{eq:dPdL-N-band}
\ee
where the cross product is defined in terms of the structure constants  as $(\partial_a \bb_n \times \bb_m)^\alpha = f_{\alpha\beta\gamma}(\partial_a b_n^\beta) b_m^\gamma $.

\subsection{Electric pseudospin polarizability \label{ssec:pseudospin-P}}

Next, we turn to the electric pseudospin responses. These are responses defined by the expectation value of an electric pseudospin, and should therefore be contrasted with the Berry phase polarization response computed from the Wannier states of a band or set of bands. An electric pseudospin is some internal microscopic degree of freedom with the symmetry of an electric dipole moment, an example of which is the layer degree of freedom associated with two-dimensional bilayer or multilayer materials. In the case of such a layer pseudospin, applying an electric field perpendicular to the layers will give rise to a charge imbalance between the layers, which can be described and calculated as an electric pseudospin polarization, in full analogy with the spin magnetization response to an external Zeeman field. 

In general, the expectation value of an observable $X$ in the ground state of the electronic system can be expressed as
\be
\langle X \rangle = \int[d\bk] \text{Tr}[\mathcal P  X] , \label{eq:<A>}
\ee
where $\mathcal P$ (i.e., the projector onto the occupied bands) is simply the ground state density matrix. The derivative of $\langle X \rangle$ with respect to $\lambda$ defines the corresponding polarizability, for which one then finds
\begin{align}
\partial_\lambda \langle X \rangle \big|_{\lambda=0} &= \int[d\bk] \text{Tr}[(\partial_\lambda \mathcal P)  X]\big|_{\lambda=0} \\
& = \int[d\bk] \text{Tr}\big[ \mathcal P^{(1)}  X\big]. \label{eq:<dA>}
\end{align}
After substituting the expression for the first order correction to the projector onto the occupied states (i.e., the density matrix), as given by Eq.~\eqref{eq:P^(1)}, we obtain the straightforward final result
\be
\partial_\lambda \langle X \rangle =  \int[d\bk] \sum_{n,m } \frac{2\text{Re}\text{Tr}\big[P_nW P_m  X\big]}{\varepsilon_n-\varepsilon_m}. \label{eq:<dA>-2}
\ee
This result for the first order correction of an expectation value is, of course, an elementary application of standard perturbation theory~\footnote{It is also directly related to the $q\rightarrow 0$ and $\omega\rightarrow$ limit of the corresponding susceptibility.}. It is applicable to any observable $X$ and in particular applies to the electric pseudospin dipole operator. For the latter we will use the symbol $\bD$ in this work. The precise form of $\bD$ depends on the specific model and microscopic context in question; an example will be discussed in Sec.~\ref{sec:2D-bilayer-AFM}. The pseudospin polarization is then defined as $\bP_D = \langle \bD \rangle $, and the corresponding polarizability is then denoted $\partial_\lambda   \bP_D   $.

As in the case of the Berry phase polarization discussed above, we now consider three classes of models and examine the form Eq.~\eqref{eq:<dA>-2} takes in terms of the parametrizations of the Hamiltonian and the perturbation. 

{\it Application to two-band models.} As in the preceding section, we start with the simplest case of a two-band model, in which case all Hermitian matrices can be expanded in the Pauli matrices $\bsigma$. In particular, the Hamiltonian can be written as $H= \bh \cdot \bsigma$, and the perturbation $W$ and the observable $X$ can be written as $W = \bw \cdot \bsigma$ and $X = \bm{x} \cdot \bsigma$, respectively. All uniform pieces (e.g. $h_0$ and $w_0$) are ignored, since they do not enter the final result. With the help of the projector onto the occupied band (see Sec.~\ref{ssec:Berry-P}) we then find that \eqref{eq:<dA>-2} reduces to
\be
\partial_\lambda \langle X \rangle = -  \int[d\bk]   \frac{ (\bh\times \bm{x})\cdot (\bh\times \bw)}{|\bh|^3}, \label{eq:dAdL-2band}
\ee
where $|\bh|$ is assumed nonzero for all $\bk$ (band gap). We thus find that the integrand of Eq.~\eqref{eq:dAdL-2band} reduces to a simple geometric expression involving the vectors $\bh$, $\bm{x}$, and $\bw$. 

{\it Application to four-band models.} Next, consider the class of four-band models constrained by $\mathcal I \mathcal T$ symmetry and defined by the unperturbed Hamiltonian $H_0 = \bh \cdot \bGamma$ (see above and Appendix~\ref{app:4-band}). Recall that here $\bh$ is a five-component real vector and $\bGamma$ are five anticommuting Hermitian matrices. To compute the polarizability for this general class of four-band models, both the perturbation $W$ and the observable $X$ should be expanded in $\Gamma$-matrices. This is achieved by writing $W = w^\alpha\Gamma_\alpha + w^{\alpha\beta}\Sigma_{\alpha\beta}$ and $X = x^\alpha\Gamma_\alpha + x^{\alpha\beta}\Sigma_{\alpha\beta}$ (with uniform pieces ignored). Both the perturbation $W$ and the observable $X$ are thus in general parametrized by a vector ($\bw$ and $\bm{x}$) and an antisymmetric tensor ($w^{\alpha\beta} $ and $x^{\alpha\beta}$, which satisfy $w^{\alpha\beta} = -w^{\beta\alpha}$ and $x^{\alpha\beta} = -x^{\beta\alpha}$).


Substituting the parametrizations of $W $ and $X$ into Eq.~\eqref{eq:<dA>-2}, as well as the projectors onto the occupied and unoccupied subspaces, we obtain
\begin{multline}
\partial_\lambda \langle X \rangle =  \int[d\bk] ~  \bigg[ \frac{2(\bh\cdot \bm{x})( \bh\cdot \bw )-2 \bh^2(\bw\cdot \bm{x}) }{|\bh|^3} \\
 -\frac{8h^\alpha h^\beta w^{\alpha\gamma}x^{\beta\gamma} }{|\bh|^3} \bigg]. \label{eq:dAdL-4band}
\end{multline}
We thus find that the polarizability can be expressed entirely in terms of $\bh$, i.e., the Hamiltonian, and the parametrizations of $W $ and $X$ given by $\bw$, $\bm{x}$, $w^{\alpha\beta}$, and $x^{\alpha\beta}$. In Sec.~\ref{ssec:bilayer-AFM-P} we consider a specific microscopic setting in which this formula applies. 

{\it Application to general $N$-band models.} Finally, we consider the general case of $N$-band models. As before, we substitute the projectors given by Eq.~\eqref{eq:SU(N)-Pn} and use an expansion of the observable $X$ and perturbation $W$ given by $X = \bm{x} \cdot \bLambda$ and $W = \bw \cdot \bLambda$, respectively. We find that the pseudospin polarizability can be written as
\begin{multline}
\partial_\lambda \langle X \rangle =  \int[d\bk] \sum_{n,m} \frac{1}{\varepsilon_{n}-\varepsilon_m} \bigg[ \frac{8}{N^2}\bw\cdot \bm{x} \\
 + \left\{ \frac{2}{N}(\bb_n+\bb_m )- \bb_n \star \bb_m \right\}\cdot (\bw\star \bm{x})   \\
 +  \frac{2}{N} \left\{( \bb_n \cdot \bm{x})(\bb_m \cdot \bw)+(\bb_m \cdot \bm{x})(\bb_n \cdot \bw) \right\} \\
 +   ( \bb_n \star \bm{x})\cdot (\bb_m \star  \bw)+(\bb_m \star  \bm{x})\cdot (\bb_n \star  \bw) \bigg]. \label{eq:dAdL-Nband}
\end{multline}
where the star product has been defined as $(\bw\star \bm{x})^\alpha = d_{\alpha\beta\gamma} w^\beta x^\gamma$, with $ d_{\alpha\beta\gamma}$ the structure constants defined in Eq.~\eqref{eq:L-anticommute} \cite{Graf:2021p085114}.

\subsection{Pseudospin polarization response to a magnetic field \label{ssec:pseudospin-P-B}}

The polarizability given by Eq.~\eqref{eq:<dA>-2} was derived using the standard perturbation theory approach outlined in Sec.~\ref{sec:problem}. It describes the response to an external field $\lambda$, which enters the Hamiltonian as in Eq.~\eqref{eq:H-Bloch}.  An external orbital magnetic field does not fall in that category, since it enters the Hamiltonian through the vector potential. This requires a more careful treatment, in particular with regard to the question of magnetic translation symmetry. If we wish to describe the pseudospin polarization response to an orbital magnetic field, we must therefore follow a different approach. 

The problem of orbital magnetic field responses was addressed in detail by Essin \emph{et al.} in Ref.~\onlinecite{Essin:2010p205104}. There, the goal was to derive microscopic expressions for the orbital magnetoelectric polarizability, which refers to the Berry phase polarization response to an orbital magnetic field. To compute the orbital magnetoelectric polarizability, Essin \emph{et al.} determined the first order correction to the ground state density matrix in the presence of a magnetic field using density-matrix perturbation theory. Here we will use this result in service of a more modest and straightforward goal: obtaining an expression for the electric pseudospin dipole response to an orbital magnetic field. Ultimately, in Sec.~\ref{sec:apply-ME}, we will demonstrate how the obtained polarizability satisfies a Maxwell relation for magnetoelectric polarizabilities associated with electric pseudospins. 

In the presence of a magnetic field, the ground state density matrix can be written as $\bar \rho = \rho^{(0)} + \rho^{(1)}$, where $ \rho^{(0)} = P$ is simply the projector onto the occupied unperturbed states and $ \rho^{(1)}$ is the correction linear in magnetic field $\bB$. Note that $\bar\rho$ here is the piece of the density matrix that has the translation symmetry of the crystal (see Ref.~\onlinecite{Essin:2010p205104}), and thus connects smoothly to the density matrix without field. The first order correction $ \rho^{(1)}$ can be decomposed into four different contributions as
\be
\rho^{(1)} = P \rho^{(1)} P+Q \rho^{(1)} Q +  P \rho^{(1)} Q +  Q\rho^{(1)} P, \label{eq:rho^1}
\ee
were we recall that $P$ and $Q$ are the projectors onto the unperturbed (i.e.,~field-free) occupied and unoccupied bands. The projection onto the subspace of occupied bands is given by
\be
P \rho^{(1)} P =   \frac{e}{4\hbar }B_a \epsilon^{abc} P F_{bc} P  , 
\ee
where $F_{bc}  =  i [ \partial_b P,\partial_cP]$ is the non-Abelian Berry curvature associated with the occupied bands. This may be seen more directly by considering the matrix elements $F^{nn'}_{bc} =  \langle u^0_n| F_{bc}| u^0_{n'} \rangle$, for which one finds 
\be
F^{nn'}_{bc} = \partial_b A^{nn'}_c-\partial_c A^{nn'}_b -i [A_b,A_c]^{nn'},
\ee
where $A^{nn'}_a = i  \langle u^0_n| \partial_a u^0_{n'} \rangle$ is the non-Abelian Berry connection associated with the unperturbed occupied bands. Similarly, the projection into the unoccupied subspace is given by 
\be
Q \rho^{(1)}Q =  - \frac{e}{4\hbar }B_a \epsilon^{abc} Q F_{bc} Q  , 
\ee
where we have used that $\partial_a Q = -\partial_aP$. The remaining terms in \eqref{eq:rho^1}, which are each other's Hermitian conjugates, connect the occupied to the unoccupied states. For $P \rho^{(1)}Q$ one has
\be
P \rho^{(1)}Q = -\frac{ie}{2\hbar }B_a \epsilon^{abc} \sum_{n,m}  \frac{P_n \{\partial_b P, \partial_c H_0 \} P_m}{\varepsilon_n-\varepsilon_m}.
\ee
Inserting all these contributions in $\langle X \rangle=\int[d\bk] \text{Tr}[\bar \rho X] $ and taking the derivative with respect to $B_a$ then yields, after some algebraic manipulations, 
\begin{multline}
\frac{\partial \langle X \rangle}{\partial B_a} = -
\frac{e}{2\hbar} \epsilon^{abc}\int[d\bk] \bigg( \text{Im}  \text{Tr} [X(P-Q)\partial_b  P \partial_c  P    ]   \\
 +\sum_{n,m} \frac{2 \text{Im}  \text{Tr} [XP_n \{ \partial_b  H_0,\partial_c  P\}P_m    ] }{\varepsilon_n-\varepsilon_m} \bigg). \label{eq:dA_dBa}
\end{multline}
This expression then describes how the expectation value of a lattice-periodic observable responds to the application of a constant uniform magnetic field. Here we will not dwell on the meaning and structure of this polarizability, in particular the two terms of the integrand, but will instead present such a discussion in the next section. The next section will show that the orbital magnetic polarizability, given in Eq.~\eqref{eq:dM_a-final}, is essentially equivalent, and we will reflect on the latter in detail. In Sec.~\ref{sec:apply-ME} we will demonstrate that the two polarizabilities are indeed connected via a Maxwell relation.

\section{Magnetic polarizability}
\label{sec:magnetic}

Following the discussion of electrical polarizability, we now focus on magnetic polarizability. Here, by magnetic polarizability we mean the magnetization response to a perturbation $W$. In full analogy to the discussion of the previous section, we consider both the orbital magnetization response and the spin magnetization response. Whereas the latter is straightforwardly determined from the ground state expectation value of spin, the former is more involved and relies on the modern theory of orbital magnetization~\cite{Thonhauser:2005p137205,Xiao:2005p137204,Ceresoli:2006p024408,Shi:2007p197202}.

\subsection{Orbital magnetization} 
 \label{ssec:orb-mag}

According to the modern theory, the orbital magnetization $\bM$ of an insulator with broken time-reversal symmetry is expressed in terms of the band energies $\varepsilon_{ n}$ and eigenstates $|  u_{ n} \rangle$ as~\cite{Thonhauser:2005p137205,Ceresoli:2006p024408}
\be
M_a = \frac{e}{2\hbar} \epsilon_{abc}\int[d\bk] \sum_{n} \text{Im} \langle \partial_b u_{ n} | (\mathcal H + \mathcal E_{ n}) | \partial_c u_{ n} \rangle. \label{eq:M_a}
\ee
(We recall that the momentum dependence of the integrand is suppressed and that $\partial_a = \partial / \partial k_a$.) While the Berry phase polarization is not gauge invariant (only polarization changes are gauge invariant), the orbital magnetization is a gauge invariant quantity. The gauge invariance of the orbital magnetization is not manifest in Eq.~\eqref{eq:M_a}, however. In order to apply perturbation theory it is necessary to start from a manifestly gauge invariant formulation of $\bM$. Such a formulation was provided in Ref.~\onlinecite{Ceresoli:2006p024408} and was achieved by rewriting the integrand of \eqref{eq:M_a} in a form which only features the Hamiltonian $H$ and the projectors onto the occupied and unoccupied subspaces $\mathcal P$ and $\mathcal Q = \mathbb{1} - \mathcal P$.  Specifically, in gauge invariant form the orbital magnetization is given by
\begin{multline}
M_a = \frac{e}{2\hbar} \epsilon_{abc}\int[d\bk]\big\{ \text{Im}  \text{Tr} [(\partial_b \mathcal P) \mathcal Q \mathcal  H\mathcal Q  (\partial_c \mathcal P)]   \\
 + \text{Im}  \text{Tr} [\mathcal  H(\partial_b \mathcal P) \mathcal Q (\partial_c \mathcal P)]  \big\}. \label{eq:M_a-gauge}
\end{multline}
It is straightforward to demonstrate that \eqref{eq:M_a-gauge} reduces to \eqref{eq:M_a} in the ``Hamiltonian gauge'' \cite{Ceresoli:2006p024408}. 

We are now in a position to determine the linear orbital magnetic polarizability, given by $\partial_\lambda M_a $, using our perturbation theory approach. We expand the projector $\mathcal P$ in the integrand of \eqref{eq:M_a-gauge} as $\mathcal P = \mathcal P^{(0)}+ \lambda \mathcal P^{(1)} =P+ \lambda \mathcal P^{(1)} $ (see Sec.~\ref{sec:problem}), do the same for $\mathcal Q$, and further make the substitution $\mathcal H = H_0 + \lambda W$. Collecting all terms linear in $\lambda$ then yields $\partial_\lambda M_a |_{\lambda=0}$, for which we find
\begin{widetext}
\begin{multline}
\partial_\lambda M_a = \frac{e}{2\hbar} \epsilon_{abc}\int[d\bk]\big\{ \text{Im}  \text{Tr} [(\partial_b P)  Q W Q  (\partial_c P)]   
 +2\text{Im}  \text{Tr} [(\partial_b \mathcal P^{(1)}) Q H_0   Q  (\partial_c  P)]+2\text{Im}  \text{Tr} [(\partial_bP)  \mathcal  Q^{(1)} H_0   Q  (\partial_c  P)] \\
  + \text{Im}  \text{Tr} [W(\partial_b  P)Q (\partial_cP)]  
 + 2\text{Im}  \text{Tr} [H_0(\partial_b \mathcal P^{(1)}) Q (\partial_c P)]+ \text{Im}  \text{Tr} [H_0(\partial_b   P) \mathcal Q^{(1)} (\partial_c  P)]  \big\}. \label{eq:dM_a}
\end{multline}
\end{widetext}
The terms involving momentum derivatives of $\mathcal P^{(1)}$ can be integrated by parts, which produces a number of additional terms. As detailed in Appendix~\ref{app:orb-mag}, most of these terms vanish, such that the final result takes the compact form
\begin{multline}
\partial_\lambda M_a =
\frac{e}{2\hbar} \epsilon_{abc}\int[d\bk] \bigg( \text{Im}  \text{Tr} [W(P-Q)\partial_b  P \partial_c  P    ]   \\
 +\sum_{n,m} \frac{2 \text{Im}  \text{Tr} [WP_n \{ \partial_b  H_0,\partial_c  P\}P_m    ] }{\varepsilon_n-\varepsilon_m} \bigg). \label{eq:dM_a-final}
\end{multline}
This expression for the orbital magnetic polarizability is the key result of this section. A strong resemblance with Eq.~\eqref{eq:dA_dBa} is immediately noticeable and will be discussed in more detail below, in particular in Sec.~\ref{ssec:ME-pseudospin}. 

A number of general observations and remarks concerning the result of Eq.~\eqref{eq:dM_a-final} can be made. The first observation is that the polarizability $\partial_\lambda M_a $ is the sum of two distinct terms. The first term only involves projectors onto the occupied and unoccupied bands (and their derivatives), but not their energies. For this reason, we refer to the first term as the \emph{geometric} term, or the geometric contribution to the polarizability. It is important to note and emphasize, however, that it does not express a purely quantum geometric property of the manifold of occupied or unoccupied states, since it also features the perturbation $W$. The latter has dimension energy, which implies that the response described by the geometric term is not quantized in units of natural constants. Hence, the geometric term does not represent a topological invariant, such as the Chern-Simons $\theta$-term of the orbital magnetoelectric polarizability in three dimensions~\cite{Qi:2008p195424,Essin:2009p146805}. A more accurate terminology might therefore be 'quasi-geometric' contribution, but here we will refer to it simply as the geometric contribution. We will show below, when we consider two- and four-band models, and also in Secs.~\ref{sec:2D-bilayer-AFM} and \ref{sec:2D-Dirac}, how the geometric term is related to, and captures, quantum geometric characteristics of the occupied states.

\begin{figure}
	\includegraphics[width=\columnwidth]{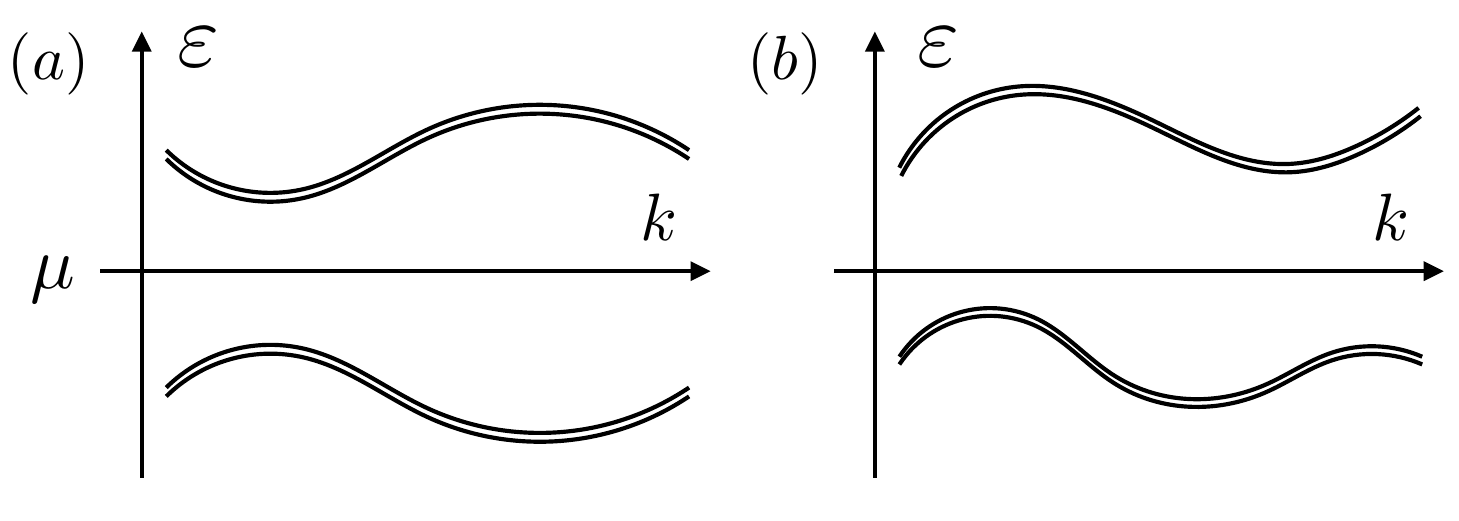}
	\caption{Schematic sketches of the energy bands of the unperturbed Hamiltonian $H_0$. (a) All valence bands and all conduction bands have the same energy $\ve^v_\bk$ and $\ve^c_\bk$ (``degeneracy''), and $\ve^v_\bk+\ve^c_\bk = \text{constant}$ (``reflection''). (b) A spectrum which has the property of degeneracy but not reflection.}
	\label{fig:interband}
\end{figure}

The second term in~\eqref{eq:dM_a-final} explicitly depends on the band energy differences via the denominator, and therefore has a more familiar perturbative structure. For this reason, we refer to this term as the \emph{interband} contribution to the polarizability. Assuming that the energy bands of the unperturbed Hamiltonian satisfy certain conditions, in particular ``degeneracy'' and ``reflection'' conditions, the interband contribution can be shown to either vanish, or reduce to a simpler form with an appealing interpretation. These conditions were first pointed out to have consequences for interband contributions in the derivation and analysis of the orbital magnetoelectric polarizability in Ref.~\onlinecite{Essin:2010p205104}. They apply similarly in the present context. Degeneracy refers to the property that, at each $\bk$, all occupied states have the same energy, similarly all unoccupied states have the same energy. Both spectra schematically sketched in Fig.~\ref{fig:interband} panels (a) and (b) have this property. Reflection refers to the additional property that the sum of valence and conduction band energies is independent of $\bk$, which is the case for the spectrum sketched in Fig.~\ref{fig:interband}(a), but not (b). If the spectrum of $H_0$ satisfies both conditions the interband contribution vanishes~\cite{Essin:2010p205104}, as is demonstrated explicitly in Appendix~\eqref{app:interband}.

Consider now the situation where the spectrum has the property of degeneracy but not reflection, as sketched in Fig.~\ref{fig:interband}(b). The occupied valence bands have energy $\ve^v$ and unoccupied conduction bands have energy $\ve^c$, but we assume that $\ve^v+\ve^c = 2h_0$ is not constant, but instead a function of $\bk$ describing a uniform piece of the dispersion (i.e., a term in the Hamiltonian proportional to the identity). As is shown in Appendix~\eqref{app:interband}, in this case the interband term reduces to
\begin{multline}
\frac{e}{\hbar} \epsilon_{abc}\int[d\bk] \sum_{n,m} \frac{\text{Im}  \text{Tr} [WP_n \{ \partial_b  H_0,\partial_c  P\}P_m    ] }{\varepsilon_n-\varepsilon_m}  \\
 \rightarrow  \frac{e}{\hbar} \epsilon_{abc}\int[d\bk] ~ h_0 \partial_b\Omega_{c\lambda}|_{\lambda=0}, \label{eq:interband-curvature}
\end{multline}
where $\Omega_{c\lambda}$ is the curvature defined in Eq.~\eqref{eq:Omega_al} and evaluated at $\lambda=0$ in (the integrand of) Eq.~\eqref{eq:dP-2}. We thus find that, under the assumption of degeneracy, the interband contribution to the orbital magnetic polarizability is an integral over the momentum derivative of the curvature $\Omega_{c\lambda}$ multiplied by the uniform dispersion $h_0$. Below we will show that the interband term indeed takes this form in the case of two-band and four-band models. In Sec.~\ref{sec:Hall}, in particular in Sec.~\ref{ssec:Berry-OMP}, we will then show that $\epsilon_{abc}\partial_b\Omega_{c\lambda}$ is equal to the Hall vector polarizability, a concept we introduce in~Sec.~\ref{sec:Hall}. 

As in the preceding section, we now consider the application of Eq.~\eqref{eq:dM_a-final} to specific classes of models. We first consider two-band and four-band models, and then turn to general $N$-band models. Examining the form Eq.~\eqref{eq:dM_a-final} takes for each of these classes provides additional insight into the nature of the geometric and interband terms.  

{\it Application to two-band models.} As in the previous section, we first consider two-band models. Here, we take the unperturbed Hamiltonian to have the form $H_0 = h_0 \mathbb{1} + \bh \cdot \bsigma$, where $h_0$ is a uniform dispersion piece proportional to the identity. This was not included in preceding sections, since it did not affect the resulting expressions for the polarizabilities. In this case, however, $h_0$ does affect the polarizability, as Eq.~\eqref{eq:interband-curvature} suggests. The form of the projectors $P$ and $Q$ onto the occupied and unoccupied band is not affected. The energies of the valence and conduction band are $\ve_{1,2}  =h_0 \mp |\bh|$, such that their difference remains $\ve_{1}  -\ve_{2}  =-2 |\bh|$. 

For the perturbation we must also include a uniform piece, denoted $w_0$, such that it takes the form $W = w_0 \mathbb{1} + \bw \cdot \bsigma$. Substituting these expressions into Eq.~\eqref{eq:dM_a-final} yields the result
\begin{multline}
\partial_\lambda M_a = 
- \frac{e}{2\hbar} \epsilon_{abc}
\int[d\bk] \bigg\{ w_0 \frac{\bh \cdot \partial_b \bh \times \partial_c \bh}{2|\bh|^3}   \\
 +(\partial_b h_0) \frac{\bw \cdot  \bh \times \partial_c \bh}{|\bh|^3}  \bigg\}, \label{eq:dM_a-2-band}
\end{multline}
where the two terms in curly brackets are the geometric term and the interband term, respectively. This expression for the orbital magnetic polarizability of two-band models has a number of properties worth highlighting, as they offer a more refined understanding of the orbital magnetic polarizability. 

The first observation is that the geometric term is equal to $w_0 \Omega_{bc}$, where $\Omega_{bc}=\bh \cdot \partial_b \bh \times \partial_c \bh/2|\bh|^3$ is the Berry curvature associated with the occupied valence band. In the case of two-band models, we thus find that the geometric term reduces to the product of a band geometric quantity---the Berry curvature---and $w_0$, the uniform piece of the perturbation, such that the contribution to the polarizability from the geometric term is an integral over the Berry curvature weighted by $w_0$. 

Similarly, the interband term also takes the form of a curvature. This becomes clear when comparing \eqref{eq:dM_a-2-band} with \eqref{eq:dPdL-2-band}, from which it follows that the interband term can be written as $ (\partial_b h_0) \Omega_{c\lambda}|_{\lambda=0} $, where $ \Omega_{c\lambda}$ is the curvature introduced in Eq.~\eqref{eq:Omega_al}. This is in full and expected agreement with the general result of Eq.~\eqref{eq:interband-curvature}. 

It is important to stress that the interpretation of the geometric and interband contributions in terms of curvatures is specific to two-band models. Such interpretation is not valid in general, as the subsequent cases of four-band and $N$-band models will show. We do however note that according to Eq.~\eqref{eq:interband-curvature} the interpretation of interband term does generalize if the degeneracy condition is satisfied. 

An interesting application of the two-band model formula of Eq.~\eqref{eq:dM_a-2-band} will be discussed in Sec.~\ref{sec:2D-Dirac}, where we consider gapped Dirac semimetals in two dimensions. This application will in particular highlight the significance of the curvatures, since the curvatures capture the nontrivial topology of Dirac points.

{\it Application to four-band models.} Let us next consider the class of four-band models introduced above. The Hamiltonian is written as $H_0 =h_0\mathbb{1} +  \bh \cdot \bGamma$ and the perturbation is again expanded as $W = w_0 \mathbb{1} +w^\alpha\Gamma_\alpha + w^{\alpha\beta}\Sigma_{\alpha\beta}$. Note that the uniform contributions $h_0$ and $w_0$ are important and enter the polarizability expressions. Using the expressions for the energies and the projectors onto the eigenspaces (see above and Appendix~\ref{app:4-band}), we find that the orbital magnetic polarizability takes the compact form
\begin{multline}
\partial_\lambda M_a = 
- \frac{e}{\hbar} \epsilon_{abc}
\int[d\bk] \bigg\{\frac{w^{\alpha\beta} h^\gamma (\partial_bh^\delta) (\partial_c h^\zeta) \epsilon_{\alpha\beta\gamma\delta\zeta}}{2|\bh|^3}   \\
 + (\partial_bh_0)\frac{w^{\alpha\beta} h^\alpha  (\partial_c h^\beta) }{|\bh|^3}  \bigg\}, \label{eq:dM_a-4-band}
\end{multline}
The first term corresponds to the geometric term and the second term corresponds to the interband term. 

Comparing Eqs.~\eqref{eq:dM_a-4-band} and \eqref{eq:dPdL-4-band} we observe that the interband term is indeed equal to $\partial_bh_0 \Omega_{c\lambda}$, as Eq.~\eqref{eq:interband-curvature} suggests. This is expected, since the four-band model manifestly satisfies the requirement of degeneracy due to $\mathcal I \mathcal T$ symmetry. The geometric term is not proportional to the Berry curvature $\Omega_{bc}$, since the Berry curvature is manifestly zero for the four-band model. 

In Sec.~\ref{sec:2D-bilayer-AFM} we will apply this four-band formula to the case of bilayer antiferromagnets in two dimensions. This discussion will clarify under what conditions the geometric term can still capture band geometric properties distinct from the Berry curvature.

{\it Application to general $N$-band model.} Finally, we consider the case of general $N$-band models. As in the previous sections, the (unperturbed) Hamiltonian takes the general form $H_0 =h_0\mathbb{1}+ \bh \cdot \bLambda$ and the perturbation is expanded as $W =w_0\mathbb{1}+ \bw \cdot \bLambda$. Since the general $N$-band case is a bit more involved, it is useful to consider the geometric term and the interband term separately. To determine the form of the geometric form we require the projectors on the occupied and unoccupied subspaces. The former is given by $P = \sum_n P_n  = N_{\text{occ}} \mathbb{1}/N  + \bb \cdot \bLambda/2$, where we have defined $\bb \equiv \sum_n \bb_n$, and the latter is simply given by $Q = \mathbb{1}- P$. With this definition we find that the geometric term can be written in terms of $\bb$, $\bw$, and $w_0$ as 
\begin{multline}
\epsilon_{abc}\int[d\bk] \text{Im}  \text{Tr} [W(P-Q)\partial_b  P \partial_c  P  ]  \\
=  \epsilon_{abc}\int[d\bk] \bigg( - w_0 \Omega_{bc} + \frac{2N_{\text{occ}}-N}{2N} \bw \cdot \partial_b\bb\times \partial_c \bb \\
+\frac12 (\bw \star \bb) \cdot \partial_b \bb  \times \partial_c \bb \bigg). \label{eq:dM_a-N-band}
\end{multline}
Here $\Omega_{bc}  = -\sum_n 2 \text{Im}   \langle  \partial_b u^0_n | \partial_c u^0_n\rangle$ is the Berry curvature of the unperturbed occupied subspace. Furthermore, as before the cross product should be understood as $(\partial_b\bb\times \partial_c \bb)^\alpha = f_{\alpha\beta\gamma}(\partial_b b^\beta) (\partial_c b^\gamma) $ and the star product is defined as $(\bw \star \bb)^\alpha = d_{\alpha\beta\gamma}w^\beta b^\gamma $~\cite{Graf:2021p085114}. This form of the orbital magnetic polarizability is insightful, as it exposes how the contributions to it relate to quantities that capture band geometric properties (i.e., the Berry curvature) or parametrize the occupied eigenspace and the perturbation (i.e., $\bb$ and $\bw$). We see from Eq.~\eqref{eq:dM_a-N-band} that there is a generic contribution proportional to the Berry curvature weighted by $w_0$.

\subsection{Spin magnetization} 
\label{ssec:spin-mag}

For completeness we next consider the spin magnetization. The analysis of the spin magnetization is very similar to that of the pseudospin polarization, since both correspond to expectation values of well-defined observables We can immediately invoke Eqs.~\eqref{eq:<A>}--\eqref{eq:<dA>-2} and apply them to physical spin $S_a$. Introducing the spin magnetization as $ M_{S,a} = g\mu_B \langle S_a \rangle/\hbar$, the spin magnetic polarizability is then simply given by
\be
\partial_\lambda M_{S,a} = \frac{g\mu_B}{\hbar} \int[d\bk] \sum_{n,m } \frac{2\text{Re}\text{Tr}\big[P_nW P_m  S_a\big]}{\varepsilon_n-\varepsilon_m}. \label{eq:<dS_a>}
\ee
Further analysis of this formula was already provided in Sec.~\ref{ssec:pseudospin-P}. All obtained expressions for two-band, four-band, or general $N$-band models apply directly to the spin magnetic polarizability. 

In Sec.~\ref{ssec:pseudospin-P-B}, we determined the pseudospin polarization response to the orbital effect of a magnetic field, which is an example of a magnetoelectric response. In the case of spin magnetization we are similarly interested in its response to an applied electric field $\bE$. As was the case with a magnetic field, an electric field does not straightforwardly fit into the perturbative treatment we have developed, and must be treated with some care. The effect of an electric field formally enters the Hamiltonian as $H_E= e \bE \cdot \br$, where $\br$ is the position operator. The latter requires careful consideration in periodic systems. The modern theory of polarization and subsequent conceptual developments provide the basis for determining the matrix elements of the position operator~\cite{Nunes:2000p155107}, which we can exploit here since we only require ``off-diagonal'' matrix elements, i.e., matrix elements between occupied and unoccupied states. 
 
To compute \eqref{eq:<dS_a>} when the electric field coupling is the perturbation, we may use that the interband matrix elements of position are given by the (interband) Berry connection~\cite{Nunes:2000p155107}:
\be
r_a^{nm} = A_a^{nm} = i \langle u^0_n | \partial_a u^0_m\rangle. \label{eq:r_a^nm}
\ee
Further using that the interband Berry connection can be expressed in terms of velocity matrix elements as $A_a^{mn} = -i \hbar v_a^{mn}/(\varepsilon_{m}-\varepsilon_n)$, with band velocity $ v_a = \hbar^{-1}\partial_a H_0$, we can substitute the matrix elements of the perturbation $W^{nm} = e r_b^{nm}  $ in \eqref{eq:<dS_a>}, such that the magnetoelectric polarizability $\partial M_{S,a} /\partial  E_b$ is obtained as
\be
\frac{\partial M_{S,a} }{\partial  E_b} = \frac{eg\mu_B}{\hbar} \int[d\bk] \sum_{n,m } \frac{2\text{Im} \big[v^{nm}_b  S^{mn}_a\big]}{(\varepsilon_n-\varepsilon_m)^2}. \label{eq:anomalous}
\ee
The spin magnetization response to an electric field has been the subject of much work in the area of spintronics, in particular in the context of current-induced spin magnetization, and (versions of) Eq.~\eqref{eq:anomalous} have appeared in a number of settings~\cite{Garate:2009p134403,Baltz:2018p015005,Manchon:2019p035004,Xiao:2022p086602,Xiao:2023p166302}. In the context of metals the integrand of the polarizability \eqref{eq:anomalous} has been referred to as ``anomalous spin''~\cite{Xiao:2023p166302}, since it can be viewed as a spin analog of the anomalous velocity~\cite{Xiao:2010p1959,Nagaosa:2010p1539}. An example of Eq.~\eqref{eq:anomalous} applied to an insulating state was discussed in Ref.~\onlinecite{Garate:2010p146802}.

\section{Hall vector polarizability \label{sec:Hall}}

The subject of this section is a type of polarizability different from the polarizabilities considered in the previous sections. Up to this point, we have examined how macroscopic bulk properties of materials, such as polarization and magnetization, respond to the application of external fields. These bulk properties correspond to densities of electric and magnetic dipole moments, and are therefore quantities defined in real space. Indeed, in general it is natural to think of polarization and magnetization as quantities with spatial dependence. 

In this section, we instead consider what we refer to as the Berry curvature or Hall vector polarizability. The Berry curvature is a momentum space density and is determined from the structure of the Bloch band eigenstates. It is the imaginary part of the quantum geometric tensor and encodes topological characteristics of a band or a manifold of bands. The Berry curvature is at the root of a number of transport phenomena of current interest (e.g., anomalous Hall effect), and it therefore natural to ask how the Berry curvature responds to external perturbations. 

Similar to polarization and magnetization, the Berry curvature is subject to symmetry constraints. The presence of inversion and time-reversal symmetry, for instance, requires the Berry curvature to vanish. Even when the Berry curvature is not forced to vanish identically, spatial or magnetic symmetries of the crystal may constrain it such that its integral over the Brillouin zone vanishes. In line with the philosophy of this paper, here we examine how the Berry curvature responds to an external perturbation which lifts (some of) the symmetry requirements which constrain it. 

Within the framework of our perturbative treatment, we therefore seek to determine how the Berry curvature changes in response to $\lambda$. The Berry curvature of the manifold of occupied bands is defined as
\be
\Omega_{ab} =  -2 \text{Im}\text{Tr}[  \mathcal P \partial_a  \mathcal P\partial_b \mathcal P ] , \label{eq:Omega_ab}
\ee
which is antisymmetric under exchange of $a$ and $b$ (i.e., $\Omega_{ab}=-\Omega_{ba}$). Integration over the Brillouin zone yields the integer-valued Chern invariant
\be
C_a = \frac{1}{4\pi}\epsilon^{abc} \int d^3\bk ~\Omega_{bc}  \label{eq:C_a}. 
\ee
When the vector $\bC$ is different from zero it describes a quantum Hall insulator in three dimensions, a weak topological phase that can be viewed as a stack of two-dimensional Chern insulators. 

Equation~\eqref{eq:C_a} suggests that $\epsilon^{abc}\Omega_{bc}/2$ is a vectorial momentum space density equivalent to $\Omega_{ab}$, and we refer to this vector as the Hall vector. The Berry curvature polarizability---which is what we are after---can therefore equivalently be referred to as the Hall vector polarizability. We note that in two dimensions the Hall vector is manifestly pinned along the third direction and becomes a pseudoscalar given by $\epsilon^{ab}\Omega_{ab}/2$, hence giving rise to a single Chern number $C$. 

Starting from Eq.~\eqref{eq:Omega_ab}, we determine the Hall vector polarizability $\partial_\lambda \Omega_{ab} $ using the same approach as in previous sections, by substituting $\mathcal P = \mathcal P^{(0)}+ \lambda \mathcal P^{(1)} =P+ \lambda \mathcal P^{(1)} $ and collecting all terms linear in $\lambda$. Before doing so, we first establish a useful relation between the polarizability $\partial_\lambda \Omega_{ab} $ and the generalized curvature $ \Omega_{a\lambda} $ introduced in Eq.~\eqref{eq:Omega_al}, which describes the change in polarization as the parameter $\lambda$ is varied. We find that the curvatures satisfy the Bianchi-type identity
\be
 \partial_\lambda \Omega_{ab}  + \partial_a \Omega_{b\lambda }  +\partial_b \Omega_{\lambda a} = 0 . \label{eq:Berry-sum}
\ee
This relation is generally valid and does not rely on perturbation theory, as is evident from the derivation presented in Appendix~\ref{app:dOmega_ab}. It follows that the Hall vector polarizability can be written as $ \partial_\lambda \Omega_{ab}   =  -\partial_a \Omega_{b\lambda }  +\partial_b \Omega_{a\lambda}$. 

It is worth reflecting on the nature of this relation. It states that the derivative of the familiar momentum space Berry curvature $\Omega_{a b}$ with respect to some parameter $\lambda$, here representing a generalized external field, is equal to momentum derivatives of the generalized curvatures $\Omega_{a\lambda}$. Hence, the Hall vector polarizability is equal to the momentum derivative of the generalized electric polarizability density, i.e., the momentum derivative of the Berry phase polarization response to the field $\lambda$. (We emphasize that, although suppressed for ease of notation, both sides of the equation are functions of $\bk$, and hence polarizability densities.)  It has become common to refer to momentum derivatives of curvatures as curvature dipoles (e.g. ``Berry curvature dipole''~\cite{Sodemann:2015p216806}), and we may therefore rephrase the interpretation of Eq.~\eqref{eq:Berry-sum} by stating that the Hall vector polarizability is equal to the generalized electric polarizability \emph{dipole}. This result, that the Hall vector polarizability is equal to an electric polarizability dipole, is the first main result of this section.

Substituting the expression for $\Omega_{a\lambda}$ found in Eq.~\eqref{eq:dP-2}, we find that the Hall vector polarizability takes the form
\be
\partial_\lambda \Omega_{ab} =  \partial_b \sum_{n,m} \frac{2 \text{Im}\text{Tr}[  P_n  ( \partial_a P_m) W]}{\varepsilon_n-\varepsilon_m} - (a\leftrightarrow b) . \label{eq:dOmega_ab} 
\ee
Based on the analysis of Sec.~\ref{ssec:Berry-P} it is straightforward to determine how this general formula applies to various classes of $N$-band models. 

An important observation is that the Hall vector polarizability is given by the total derivative of a gauge invariant quantity. This has the implication that the integral of the Hall vector polarizability over the full Brillouin zone vanishes---an eminently sensible result, since the Chern invariant given by \eqref{eq:C_a} should not change in response to a small perturbation. Indeed, the topological transition required to change the Chern invariant is manifestly beyond the reach of a perturbative approach. 

In light of this, let us briefly review the bidding. We are interested in determining changes of the Berry curvature (or Hall vector) given by Eq.~\eqref{eq:Omega_ab}, which is a $\bk$-dependent quantity (i.e., a momentum space density) associated with the manifold of occupied bands of an insulator. When integrated over the full Brillouin zone, the Berry curvature yields the integer-valued Chern invariant. The Hall vector \emph{polarizability}, given by Eq.~\eqref{eq:dOmega_ab}, is also a momentum space density and describes the change of the Berry curvature at given momentum $\bk$. It is as such that the polarizability is of interest. When integrated over the Brillouin zone it must yield zero, which means that it is not of immediate consequence for insulators. In a more general setting, beyond the specific context of insulators, the Hall vector polarizability can be of great physical significance, since in metals one may be interested in responses which require integration over only part of the Brillouin zone (i.e., the occupied part). This motivates further study of the Hall vector polarizability. In Sec.~\ref{sec:apply-altermagnet} we demonstrate how studying the Hall vector polarizability can give valuable insight into the response behavior of a simple model for an altermagnet.  

A final remark concerns the fact that the Berry curvature, as defined by Eq.~\eqref{eq:Omega_ab}, is the Berry curvature associated with the entire manifold of occupied bands. In a multi-band system it may therefore be the sum of contributions from individual bands. It is straightforward to generalize the above analysis to determine the contribution from an individual band. In fact, one may simply interpret the projector appearing Eq.~\eqref{eq:Omega_ab} as the projector onto a single (isolated) band. The only difference is then that in Eq.~\eqref{eq:dOmega_ab} the sum over $n$ includes only a single term---the band of interest---and the sum over $m$ should be read as a sum over all other bands.

\subsection{Relationship to orbital magnetic polarizability}
\label{ssec:Berry-OMP}

The relation given by Eq.~\eqref{eq:Berry-sum} can be used to gain further insight into the expressions obtained for the orbital magnetic polarizability in Sec.~\ref{ssec:orb-mag}. Specifically, the relation between the Hall vector polarizability and the curvature dipoles allows us to recast the interband term given by Eq.~\eqref{eq:interband-curvature} in terms of the Hall vector polarizability. Recall that Eq.~\eqref{eq:interband-curvature} gives the form of the interband contribution when the unperturbed spectrum satisfies the property of degeneracy (see Fig.~\ref{fig:interband}). With this assumption, and given the identity \eqref{eq:Berry-sum}, the interband term can be rewritten as
\be
 \epsilon_{abc}\int[d\bk] ~ h_0 \partial_b\Omega_{c\lambda}  =  -\frac{\epsilon_{abc}}{2}\int[d\bk] ~ h_0 \partial_\lambda\Omega_{bc} , \label{eq:interband-Hall-vector}
\ee
where it is understood that both $\partial_b\Omega_{c\lambda}$ and $\partial_\lambda\Omega_{bc}$ are evaluated at $\lambda=0$. We thus find that the interband contribution to the orbital magnetic polarizability reduces to an integral over the Hall vector polarizability, multiplied by $h_0$. This shows that the Hall vector polarizability does in fact enter the expression for the (orbital) magnetic response of an insulator.

\subsection{Two-band and four-band models}
\label{ssec:Berry-2-4band}

Let us next consider the expression for the Hall vector polarizability obtained for the general two-band and four-band models introduced in previous sections. Since the curvature $\Omega_{a\lambda}$ takes a simple form in these cases, as seen in Eqs.~\eqref{eq:dPdL-2-band} and \eqref{eq:dPdL-4-band}, the Hall vector polarizability can similarly be expressed in terms of the parametrization of the Hamiltonian and the perturbation. We will not revisit the definitions of the two- and four-band models, but simply state the results and refer the reader to Sec.~\ref{sec:electric} (as well as Sec.~\ref{sec:magnetic}) for details. 

In the case of two-band models, and after some straightforward manipulations, the Hall vector polarizability reduces to the expression
\begin{multline}
\frac12\epsilon_{abc} \partial_\lambda\Omega_{bc} =  \epsilon_{abc} \left\{ \frac{ \partial_b(\bh \times  \bw ) \cdot  \partial_c \bh}{2|\bh|^3}  \right. \\
\left. + \frac{3\bh \times( \partial_b \bh \times \partial_c \bh ) \cdot \bh\times \bw}{4|\bh|^5} \right\}. \label{eq:dOmega-2band}
\end{multline}
This shows that the polarizability can be written entirely in terms of vectorial functions ($\bh $ and $\bw$) which parametrize the Hamiltonian and the perturbation.

In the case of four-band models, we find the Hall vector polarizability as
\begin{multline}
\frac12\epsilon_{abc} \partial_\lambda\Omega_{bc} = 2\epsilon_{abc} \left\{ \frac{  (\partial_bh^\beta) ( \partial_c h^\alpha) w^{\alpha\beta} }{|\bh|^3 } + \right. \\
\left. \frac{ h^\beta ( \partial_c h^\alpha) (\partial_b w^{\alpha\beta} )}{|\bh|^3 }-  \frac{3 h^\beta( \partial_c h^\alpha) w^{\alpha\beta} \bh \cdot \partial_b\bh}{|\bh|^5} \right\},  \label{eq:dOmega-4band}
\end{multline}
which is simply the momentum derivative of the integrand of Eq.~\eqref{eq:dPdL-4-band}. The Hall vector polarizability of generic $N$-band models can similarly be obtained by noting that $\Omega_{c\lambda}$ is equal to the integrand of Eq.~\eqref{eq:dPdL-N-band} and taking the appropriate derivative.

\section{Magnetoelectric polarizabilities and Maxwell relations }
\label{sec:apply-ME}

In the preceding sections we have derived expressions for a variety of different polarizabilities. In this section we take a first step towards applying these general results. Our specific goal in this section is to examine magnetoelectric polarizabilities, i.e., polarizabilities which reflect the cross-coupling of magnetic and electric degrees of freedom, and demonstrate from a microscopic point of view that distinct types of magnetoelectric polarizabilities satisfy the required Maxwell relation. Such Maxwell relations are expected based on general arguments, and therefore not surprising, but it is nonetheless worthwhile to demonstrate explicitly that the obtained microscopic expressions for the polarizabilities are indeed equal. Not in the least since the expected Maxwell relations are not obvious from the microscopic starting points. 

One such magnetoelectric Maxwell relation was established previously in Refs.~\onlinecite{Essin:2010p205104} and \onlinecite{Malashevich:2010p053032}. Both works set out to obtain microscopic expressions for the orbital magnetoelectric polarizability, but adopted different yet complementary approaches. Whereas Ref.~\onlinecite{Essin:2010p205104} determined the (Berry phase) polarization response to an orbital magnetic field of insulators in three dimensions, Ref.~\onlinecite{Malashevich:2010p053032} determined the orbital magnetization response to an electric field. Comparing the resulting expressions for the orbital magnetoelectric polarizability shows that both approaches yield the same result, thus establishing that the Maxwell relation
\be
\frac{\partial P_a}{\partial B_b} = \frac{\partial M_b}{\partial E_a} , \label{eq:maxwell-orb}
\ee
indeed holds. Here $\bP$ and $\bM$ are the Berry phase polarization and the orbital magnetization, respectively, and $\bB$ and $\bE$ are the magnetic and electric fields, which enter the Hamiltonian through the vector potential ($\bB = \boldsymbol{\nabla} \times \bA$) and as a term $e\bE \cdot \br$. 

In the remainder of this section, we instead focus on the spin magnetoelectric polarizability and the electric dipole magnetoelectric polarizability. By the former we mean the polarization response to a Zeeman magnetic field and by the latter we mean the orbital magnetization response to an electric displacement field. Since the Zeeman field couples to spin, the polarization response is referred to as the \emph{spin} magnetoelectric polarizability, and similarly, since the electric displacement field couples to an electric dipole pseudospin, the magnetization response is referred to as the \emph{electric dipole} magnetoeletric polarizability. 

\subsection{Magnetoelectric responses and spin \label{ssec:ME-spin}}

Consider first the spin magnetoelectric polarizability, the polarization response to a magnetic field $\bB_S$. Here we add the subscript $S$ (spin) to emphasize that we are interested in the response generated by the coupling of the magnetic field to the electron spin. This coupling is given by the Zeeman term in the Hamiltonian and takes the form
\be
H_S = -  \frac{g\mu_B}{\hbar}  \bS \cdot \bB_S , \label{eq:H_S}
\ee
where $g$ is the gyromagnetic ratio and $\mu_B$ is the Bohr magneton. The Zeeman term should therefore be regarded as the perturbation in Eq.~\eqref{eq:H-Bloch} and $\bB_S$ corresponds to the external field. To compute the magnetoelectric polarizability $\partial P_a/ \partial B_{S,b}$, we can invoke Eq.~\eqref{eq:dP-3}, the general formula for the Berry phase polarization response, and directly find
\be
\frac{\partial P_a}{\partial B_{S,b}} = \frac{eg\mu_B}{\hbar}  \int [d\bk] \sum_{n,m}  \frac{2\text{Im}[ v_a^{nm} S_b^{mn}]}{(\varepsilon_{n}-\varepsilon_m)^2}. 
\ee
This expression, which was first obtained in this form in Ref.~\onlinecite{Gao:2018p134423}, is equal to the spin magnetization response ($M_{S,b}$) to an electric field ($E_a$) given by Eq.~\eqref{eq:anomalous}. We therefore microscopically establish the Maxwell relation
\be
\frac{\partial P_a}{\partial B_{S,b}} = \frac{\partial M_{S,b}}{\partial E_a} .   \label{eq:Maxwell-spin}
\ee

\subsection{Magnetoelectric responses and electric dipole pseudospin \label{ssec:ME-pseudospin}}

Next, we focus on the electric dipole magnetoelectric polarizability. This magnetoelectric polarizability refers to the orbital magnetization response to an electric field, caused by the Zeeman-like coupling to the electric pseudospin, and similarly the pseudospin polarization response to an orbital magnetic field. Therefore, the electric dipole pseudospin plays the key role here, precisely the role played by spin in case of the spin magnetoelectric polarizability. Our aim here is simply to highlight that, as far as the microscopic expressions for the polarizabilities are concerned, these responses are indeed the same.

The electric field couples to the electric dipole pseudospin $\bD$ in a way that is analogous to the Zeeman coupling. The term in the Hamiltonian is written as
\be
H_D = - \bD \cdot \bE_D ,\label{eq:H_D}
\ee
where we have added the subscript $D$ to the electric field, to emphasize its coupling to the pseudospin $\bD$. The precise form of $\bD$ will depend on the specific context in question. A setting in which an electric pseudospin naturally arises is the field of layered two-dimensional materials. The layer degree of freedom realizes an intrinsic electric pseudospin and couples to a perpendicularly applied electric field as described by \eqref{eq:H_D}. In Sec.~\ref{sec:2D-bilayer-AFM}, we consider an example of such a pseudospin in the context of bilayer magnets. 

Consider the orbital magnetization response to an electric field which couples to the pseudospin. Such a response is described by Eq.~\eqref{eq:dM_a-final} with $W = - D_b$ ($b=x,y,z$), as per Eq.~\eqref{eq:H_D}. Similarly, the pseudospin polarization response to an orbital magnetic field is described by Eq.~\eqref{eq:dA_dBa}, with $X= D_b$ and $\langle X \rangle =  P_{D,b}$. Comparing the two resulting expressions for the polarizabilities, it immediately follows that
\be
\frac{\partial M_a}{\partial E_{D,b}} =\frac{\partial P_{D,b}}{\partial B_a} , \label{eq:Maxwell-pseudospin}
\ee
which establishes the microscopic Maxwell relation.

\section{Application to antiferromagnet in 1D}
\label{sec:apply-1D}

In this section we apply the general results of the previous sections to a toy model antiferromagnet in one dimension. Our goal is to examine the magnetoelectric properties of this toy model antiferromagnet by computing the (spin) magnetoeletric polarizability using the general polarizability formulas. 

\begin{figure}
	\includegraphics[width=\columnwidth]{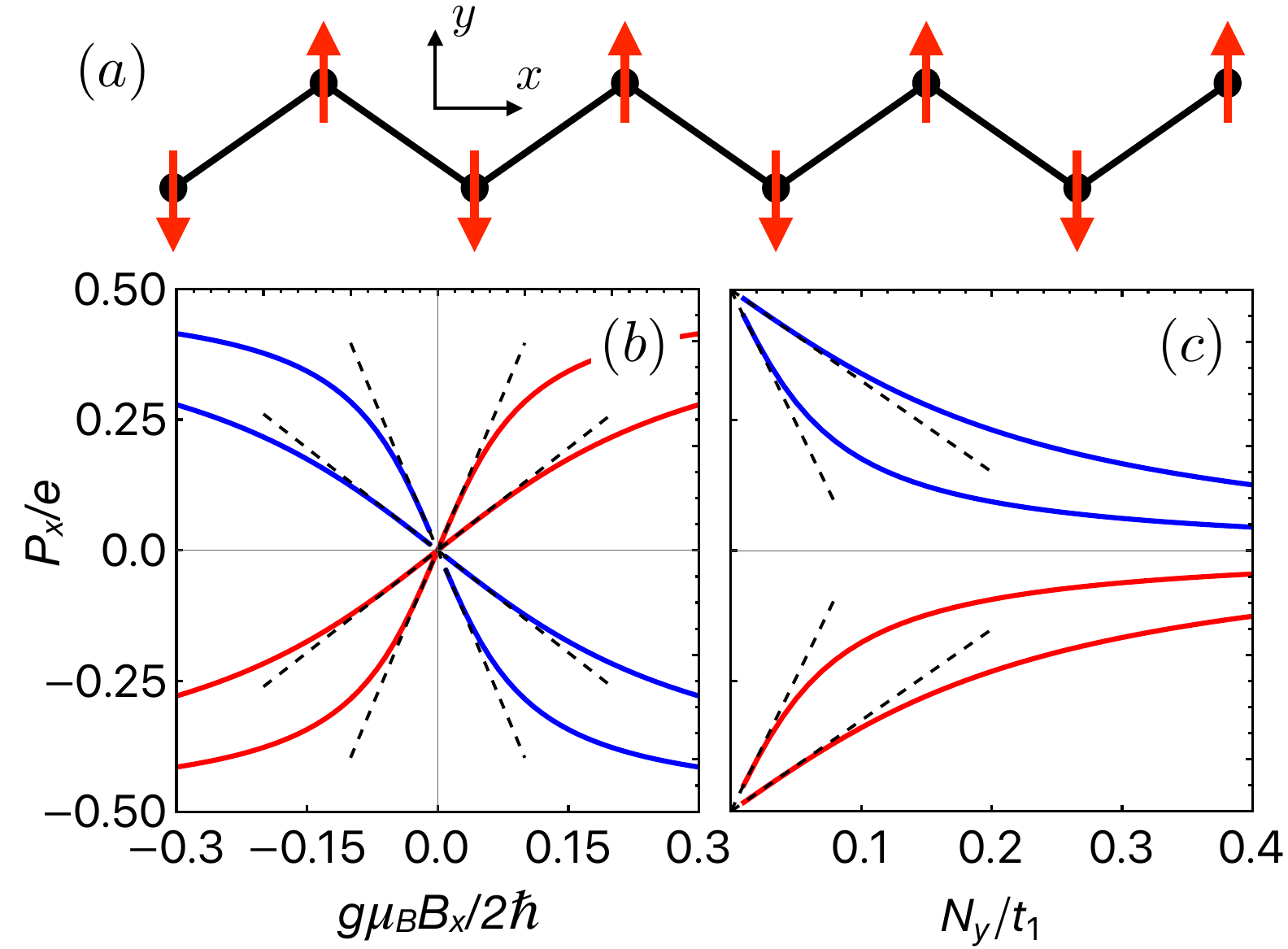}
	\caption{(a) Depiction of the N\'eel antiferromagnetic zigzag chain in 1D. Red arrows indicate the ordered moments, which are chosen to point in the $\hat y$ direction. (b) Calculation of the Berry phase polarization $P_x$ as a function of Zeeman field $B_x$ for the 1D antiferromagnetic chain model, as defined in Eqs.~\eqref{eq:H_0-1D-A} and \eqref{eq:W-1D-A}. Red and blue curves correspond to $N_y/t_1 = 0.05, 0.15$ and $N_y/t_1 = -0.05, -0.15$, respectively. (We have set $t_{\text{SO}} = 0.8t_1$.) The dashed straight lines correspond to the linear polarizabilities $\partial P_x /\partial B_x$ computed using Eq.~\eqref{eq:dPx/dBx}. (c) Calculation of the Berry phase polarization $P_x$ as a function of N\'eel order parameter $N_y$ in the presence of a finite Zeeman field $B_x$. Here red and blue curves correspond to $b_x/t_1 = 0.1, 0.3$ and $b_x/t_1 = -0.1, -0.3$, respectively. The dashed straight lines correspond to the linear polarizabilities $\partial P_x /\partial N_y$ computed using Eq.~\eqref{eq:dPx/dNy}. As $N_y\rightarrow 0 $ the polarization $P_x/e$ tends to the value $-\frac12 \text{sgn}(b_x)$, which reflects the topological half-quantized polarization of the inversion symmetric insulating state at $N_y=0$. }
	\label{fig:1D_AFM}
\end{figure}

The specific model we study is the one-dimensional zigzag chain shown in Fig.~\ref{fig:1D_AFM}(a), which was considered in prior work as a useful idealized model for magnetoelectric coupling in experimentally relevant antiferromagnets~\cite{Yanase:2014p014703,Hayami:2015p064717,Hayami:2022p123701,Yatsushiro:2022p155157,Hayami:2023p094106}. While the zigzag chain does have inversion symmetry ($\mathcal I$), the magnetic sites are not themselves  inversion centers, and this allows for a linear magnetoelectric response in the N\'eel state. Collinear N\'eel order is depicted in Fig.~\ref{fig:1D_AFM}(a) by red arrows. It indeed breaks the inversion symmetry of the zigzag chain but preserves the product of time-reversal and inversion symmetry ($\mathcal I \mathcal T$).  

To construct an electronic model for the zigzag antiferromagnet, we introduce the electron destruction operators $c_{\bk \alpha \sigma}$, where $\alpha=A,B$ labels the two sublattices of the chain and $\sigma=\up,\down$ denotes spin. The Hamiltonian is expressed as
\be
H = \sum_{\bk} c^\dagger_\bk \mathcal H_\bk c_\bk, \;\; c_{\bk} = ( c_{\bk A\up},c_{\bk A\down},c_{\bk B\up},c_{\bk B\down})^T, \label{eq:H_k-4band}
\ee
where $\mathcal H_\bk$ is of the form $\mathcal H_\bk = H_{0,\bk} + \lambda W$, as in Eq.~\eqref{eq:H-Bloch}. In what follows we take the N\'eel vector in the $\hat y$ direction, as shown in Fig.~\ref{fig:1D_AFM}(a), in which case straightforward symmetry arguments predict a polarization $P_x$ in response to a Zeeman field $B_{S,x}$ applied in the $\hat x$ direction. We seek to compute the associated polarizability using methods discussed in the previous sections (and in previous works \cite{Gao:2018p134423,Venderbos:arXiv2025}). A minimal model for the antiferromagnet without Zeeman field takes the form \cite{Yanase:2014p014703}
\be
H_0 = -2t_1 \cos \tfrac{k}{2}\tau^x +  t_{\text{SO}} \sin k ~ \tau^z\sigma^z +N_y \tau^z\sigma^y, \label{eq:H_0-1D-A}
\ee
where $\btau = (\tau^x,\tau^y,\tau^z)$ and $\bsigma = (\sigma^x,\sigma^y,\sigma^z)$ are Pauli matrices corresponding to the sublattice and spin degree of freedom, respectively. Here $t_1$ is the nearest neighbor hopping amplitude, $t_{\text{SO}}$ is the strength of spin-orbit coupling, and $N_y$ is the N\'eel order parameter. The Zeeman term is taken as the perturbation, and we write it as 
\be
\lambda W = b_x \sigma^x, \qquad b_x \equiv - \frac{g \mu_B}{2\hbar }B_x , \label{eq:W-1D-A}
\ee
where $b_x$ has been introduced as an effective Zeeman field with dimension energy. It is straightforward to verify that $H_0$ given by \eqref{eq:H_0-1D-A} indeed has a gapped (and, in fact, twofold degenerate) spectrum, such that the half-filled system is an insulator. 

To compute the magnetoelectric polarizability, we make use of the results presented in Sec.~\ref{ssec:Berry-P}, in particular the formulas applicable to two-band models. The two-band model formula of Eq.~\eqref{eq:dPdL-2-band} can be applied after block diagonalizing the full four-band Hamiltonian into two $2\times 2$ blocks, which is achieved by exploiting the twofold rotation symmetry $\{ C_{2x} | \tfrac12 \}$ (i.e., rotation followed by half-translation) of the antiferromagnetic zigzag chain. Even in the presence of a Zeeman field along $\hat x$, the 1D chain remains invariant under the twofold rotation $C_{2x}$ followed by half translation. This symmetry implies that $\mathcal H_k $ commutes with $\tau^x\sigma^x$ and can therefore be brought in block diagonal form by expressing it in a basis of eigenstates of $\tau^x\sigma^x$. To make contact with the notation used in Sec.~\ref{ssec:Berry-P}, we parametrize $H_0$ as $H_0 = h^x \tau^x + h^y \tau^z\sigma^y+ h^z \tau^z\sigma^z  $, with $(h^x,h^y,h^z) = (-2t_1 \cos \tfrac{k}{2},N_y, t_{\text{SO}} \sin k)$, and obtain the two diagonal blocks as
\be
\mathcal H^\pm_k =   \begin{pmatrix} h^z &  \pm h^x  -ih^y \\ \pm h^x  +ih^y & -h^z  \end{pmatrix} +    \begin{pmatrix} 0&  b_x \\ b_x & 0  \end{pmatrix}.
\ee
Here $\pm$ corresponds to the eigenvalues $\pm 1$ of $\tau^x\sigma^x$. The Hamiltonian within each block now has the form discussed in Sec.~\ref{ssec:Berry-P} and we can apply Eq.~\eqref{eq:dPdL-2-band} to determine the polarizability $\partial P_x /\partial B_x$. The total polarizability is obtained as the sum of the two blocks and we find
\begin{align}
\frac{\partial P_x}{\partial B_x} & = -\frac{e g\mu_B}{2\hbar }  \int \frac{dk}{2\pi}   \frac{h^y \partial_k h^z }{(\bh\cdot \bh)^{3/2}}, \\
& = -\frac{e g\mu_B}{2\hbar }  \int \frac{dk}{2\pi}    \frac{t_{\text{SO}} N_yc_k }{(N^2_y+t_{\text{SO}}^2s^2_k+4t_1c^2_{k/2})^{3/2}}.\label{eq:dPx/dBx}
\end{align}
Here we have abbreviated $c_{k}\equiv \cos k$, $s_{k}\equiv \sin k$, and $c_{k/2}\equiv \cos k/2$. 

In Fig.~\ref{fig:1D_AFM}(b) we show the Berry phase polarization $P_x$ as a function of Zeeman field $B_x$ for the 1D chain model defined by Eqs.~\eqref{eq:H_0-1D-A} and \eqref{eq:W-1D-A}, calculated using modern theory methods~\cite{Resta:1992p51,King-Smith:1992p1651,Vanderbilt:1993p4442,Resta:1994p899}. Different curves are obtained for different values of the N\'eel order parameter $N_y$, with red (blue) curves corresponding to $N_y>0$ ($N_y<0$). As expected, the polarization vanishes when $B_x=0$, since the product of inversion and time-reversal forbids a nonzero polarization. Furthermore, the polarization response to $B_x$ is stronger for larger $N_y$---also as expected. 

The main purpose of computing the polarization $P_x$ is to demonstrate that the polarizability formula \eqref{eq:dPx/dBx} correctly captures the response to $B_x$ in the linear regime. The dashed straight lines in Fig.~\ref{fig:1D_AFM}(b) indicate the polarization calculated via the relation $P_x =(\partial P_x /\partial B_x) B_x$, with $\partial P_x /\partial B_x$ given by \eqref{eq:dPx/dBx}. Comparison with the direct calculation of $P_x$ confirms that the polarizability formula correctly describes the linear response to a Zeeman field $B_x$.

Another possible application of the linear polarizability formalism is to regard the N\'eel order term as the perturbation. In this case, the unperturbed Hamiltonian of the zigzag chain in the presence of the Zeeman field is given by 
\be
H_0 = -2t_1 \cos \tfrac{k}{2}\tau^x +  t_{\text{SO}} \sin k ~ \tau^z\sigma^z + b_x \sigma^x.  \label{eq:H_0-1D-B}
\ee
This Hamiltonian also describes an insulator at half filling, but the two occupied valence bands are not manifestly degenerate, which is a result of broken $\mathcal I \mathcal T$ symmetry. The Zeeman field clearly breaks $\mathcal T$ but preserves inversion symmetry $\mathcal I$, and $H_0$ therefore describes an inversion-symmetric insulator. N\'eel order can then be introduced as a perturbation, as given by
\be
\lambda W =N_y \tau^z\sigma^y, \label{eq:W-1D-B}
\ee
and its effect is to break inversion symmetry such that a nonzero polarization $P_x$ is now allowed. 

In Fig.~\ref{fig:1D_AFM}(c) we show $P_x$ as a function of $N_y$, for four different values of the effective Zeeman field $b_x$ (red and blue curves). An important feature is immediately noticeable: as $N_y \rightarrow 0 $ the polarization does not tend to zero, but instead tends to the half-quantized value $\pm 1/2$. This is due to the fact that the half-filled insulator at $N_y = 0 $ is a nontrivial topological crystalline phase protected by inversion symmetry~\cite{Hughes:2011p245132}. Inversion-symmetric insulators in one dimension admit a $\mathbb{Z}_2$ classification and the distinction between the two classes is reflected in the electric polarization. Since inversion symmetry requires that $P_x = -P_x$ up to an integer multiple of $e$, the two admissible values of polarization are $P_x/e = 0,1/2$ (modulo an integer). The spin-orbit coupled zigzag chain in the presence of a Zeeman field realizes an insulator (at a filling of two electrons per unit cell) with topological half-quantized polarization~\cite{Venderbos:arXiv2025}. Note that the curves for $b_x>0$ tend to $P_x = e/2$ as $N_y \rightarrow 0 $, whereas the curves for $b_x<0$ tend to $P_x = -e/2$. Precisely at $N_y=0$ there is no distinction, since polarization is only defined up to an integer multiple of $e$, but $N_y$ acts as a symmetry-breaking term which fixes the sign of the polarization, as clearly evidenced by Fig.~\ref{fig:1D_AFM}(c). 

While topology sets the value of $P_x$ at $N_y=0$, the response of $P_x$ to nonzero $N_y$ is given by the linear polarizability $\partial P_x/ \partial N_y$, which is calculated in general using Eq.~\eqref{eq:dP-2}, but in this case can be computed using Eq.~\eqref{eq:dPdL-2-band}. As per our argument above, in the presence of N\'eel order $N_y$ and a Zeeman field $B_x$, the Hamiltonian of the zigzag chain decomposes into two two-band Hamiltonians, which allows the use of Eq.~\eqref{eq:dPdL-2-band}. Adding the contributions of the two occupied valence bands yields
\be
\frac{\partial P_x}{\partial N_y} = e \sum_{\eta =\pm} \int \frac{dk}{2\pi}    \frac{\eta 2 t_{\text{SO}} t_1 c^3_{k/2} -b_x t_{\text{SO}} c_k  }{2[t_{\text{SO}} ^2s^2_k+(b_x -\eta 2t_1c_{k/2})^2]^{3/2}}, \label{eq:dPx/dNy}
\ee
where we note that in the presence of the Zeeman field the valence bands are nondegenerate ($\mathcal I \mathcal T$ is broken). 

The polarization obtained from Eq.~\eqref{eq:dPx/dNy} is shown in Fig.~\ref{fig:1D_AFM}(c) by the black dashed lines, showing perfect agreement with the full Berry phase calculation of $P_x$ in the linear response regime.

\section{Application to bilayer antiferromagnets in 2D}
\label{sec:2D-bilayer-AFM}

We continue our survey of applications by turning to antiferromagnets in two dimensions (2D). More specifically, this section will focus on a class of bilayer N\'eel antiferromagnets composed of two anti-aligned ferromagnetic layers. Depending on the orientation of the N\'eel vector such bilayer antiferromagnets can exhibit magnetoelectric responses associated with the layer degree of freedom~\cite{Gong:2013p2053,He:2020p1650,Tao:2024p096803,Fan:2024p7997,Hu:2025arXiv,Venderbos:arXiv2025,Radhakrishnan:arXiv2025}.  An example is the orbital magnetization response to a perpendicular electric displacement field, which we will describe and discuss below based on the results of Sec.~\ref{ssec:orb-mag}. We also discuss a more straightforward response to an applied perpendicular displacement field, which is the electric pseudospin polarization. 

A microscopic model for the bilayer system can be cast in the same form as Eq.~\eqref{eq:H_k-4band}, where $A$ and $B$ should now be interpreted as a label for the two layers. The Hamiltonian $\mathcal H_\bk  $ is thus a four-band Hamiltonian resulting from the layer pseudospin ($A,B$) and physical spin ($\up,\down$) degrees of freedom. The unperturbed Hamiltonian $H_{0,\bk}$ takes the general form~\cite{Venderbos:arXiv2025,Radhakrishnan:arXiv2025}
\be
H_{0,\bk} =\varepsilon_{0,\bk} +   f_{x,\bk} \tau^x+ f_{y,\bk} \tau^y+  \tau^z \bn_\bk \cdot \bsigma , \label{eq:H_k-2D}
\ee
where $f_{x/y,\bk}$ and $\bn_\bk$ are momentum-dependent functions parametrizing the Hamiltonian. The form of $H_{0,\bk}$ ensures that it is compatible with a manifest symmetry of the antiferromagnetic bilayer: the product of spatial inversion ($\mathcal I$) and time-reversal ($\mathcal T$) symmetry. Antiferromagnetism breaks both inversion symmetry, which involves the exchange of layers, and time-reversal symmetry, but the bilayer antiferromagnet remains invariant under $\mathcal I \mathcal T$. (See the discussion of the zigzag chain of the previous section.) Additional crystalline symmetries can impose further constraints on \eqref{eq:H_k-2D}, but we will focus on its most general form, constrained only by $\mathcal I \mathcal T$. The upshot of the $\mathcal I \mathcal T$ symmetry requirement is that Eq.~\eqref{eq:H_k-2D} falls in the special class of four-band models discussed in Sec.~\ref{sec:electric}. The term $\varepsilon_{0,\bk}$ in $H_{0,\bk} $ describes a uniform dispersion proportional to the identity (i.e., what we had more generally denoted $h_0$ in previous sections). 

In what follows we shall assume that the N\'eel vector points out of the plane (i.e., easy-axis anisotropy). This implies that N\'eel order enters as $N_z \tau^z \sigma^z$ in the Hamiltonian, and is therefore part of $n^z_\bk$. The underlying assumption of an insulator requires that  $(f^2_{x,\bk} + f^2_{y,\bk} + \bn^2_\bk)^{1/2}$ never vanishes for any $\bk$, which is guaranteed when $n_{z,\bk}  $ is simply given by $n_{z,\bk} = N_z$. We further tacitly assume the absence of an indirect band gap closing. 

Since we are interested in determining the response of the bilayer to a the perpendicular displacement field $E_{D,z}$, consider a perturbation of the general type of Eq.~\eqref{eq:H_D}. In the present context, the $\hat z$ component of the electric dipole operator takes the form $D_z = -ed\tau^z$, where $d$ is the distance between the layers, such that the perturbation describing the effect of the displacement field reads as
\be
 H_D= ed E_{D,z} \tau^z . \label{eq:E_Dz}
\ee
The full Hamiltonian is then $\mathcal H = H_0 + H_D$, and the polarizabilities of interest can be computed using the machinery of the previous sections.

\subsection{Electric dipole pseudospin response to perpendicular electric field}
\label{ssec:bilayer-AFM-P}

Consider first the electric pseudospin response to the perpendicular displacement field $E_{D,z} $. Such a response is described by the general polarizability formula of Eq.~\eqref{eq:<dA>-2} with $X = D_z = -ed \tau^z$ for the bilayer electric dipole moment and $W = ed \tau^z$ for the perturbation (with $\lambda = E_{D,z}$). We seek to determine the polarizability $\partial P_{D,z}/ \partial E_{D,z}$, with $P_{D,z} = \langle D_z \rangle $. Since the Hamiltonian of Eq.~\eqref{eq:H_k-2D} falls in the special class of four-band models discussed in Sec.~\ref{ssec:pseudospin-P}, computing the polarizability is a straightforward application of the formula obtained in  Eq.~\eqref{eq:dAdL-4band}. In the present case, the specific representation of $\Gamma$-matrices takes the form $(\Gamma_1,\Gamma_2,\Gamma_{3,4,5}) = (\tau^x,\tau^y,\tau^z\bsigma)$, which is directly inferred from \eqref{eq:H_k-2D}. (More details on this representation are given in Appendix \ref{app:4-band}.) With this identification of $\Gamma$-matrices the observable $X$ is expressed as $X = x^{12}\Sigma_{12} + x^{21}\Sigma_{21} $, with $x^{12}=-a^{21} = -ed/2$, and the perturbation is written as $W = w^{12}\Sigma_{12} + w^{21}\Sigma_{21} $, with $w^{12}=-w^{21} = ed/2$. Straightforward application of Eq.~\eqref{eq:dAdL-4band} then yields the result
\be
\frac{\partial P_{D,z}}{\partial E_{D,z}} = -e^2d^2 \int \frac{d^2\bk}{(2\pi)^2} \frac{2f^2_{\bk}}{(f^2_{\bk}+\bn_\bk^2)^{3/2}}, \label{eq:dPz/dEz}
\ee
where we have defined $f^2_{\bk} \equiv f^2_{x,\bk}+f^2_{y,\bk}$. We thus find that the polarizability is expressed simply in terms of the parametrization of the Hamiltonian. Note that it is not surprising that the numerator in Eq.~\eqref{eq:dPz/dEz} is given by $f^2_{\bk}$, since $f_{x,\bk}$ and $f_{y,\bk}$ are the coefficients of $\tau^x$ and $\tau^y$ in \eqref{eq:H_k-2D}, respectively, and therefore determine the extent of hybridization between the layers; when $f_{x,\bk}= f_{y,\bk}=0$ there is no coupling between the layers, causing the polarizability to vanish. It is further worth observing that the electric polarizability given by Eq.~\eqref{eq:dPz/dEz}, which, as far as symmetry is concerned, describes a conventional polarization response to an electric field, does not rely on antiferromagnetic nature of the bilayer. Since the N\'eel vector enters the Hamiltonian via $\bn_\bk$, and not via $f_{x,\bk}$ or $f_{y,\bk}$, Eq.~\eqref{eq:dPz/dEz} need not vanish in the nonmagnetic state, consistent with elementary symmetry arguments. 

As a concrete realization of the general model given by Eq.~\eqref{eq:H_k-2D}, consider the surface state model of a quasi-2D magnetic topological insulator thin film~\cite{Otrokov:2017p025082,Otrokov:2019p416,Rienks:2019p423,Li:2019peaaw5685,Zhang:2019p206401,Liu:2020p522,Lee:2019p012011,Lei:2020p27224,Deng:2020p895,Deng:2021p36,Gao:2021p521}. Such a model consists of two surface Dirac fermions, one for the top and one for the bottom surface, which are coupled by a tunneling amplitude $t_\perp$. In this specific model $\bn$ is given by $\bn = (\hbar v k_y, -\hbar v k_x, N_z)$, where $v$ is the Fermi velocity of the surface Dirac states, and $f_x = t_\perp$. Inserting this into Eq.~\eqref{eq:dPz/dEz} and taking the integral gives the result  $\partial P_{D,z}/ \partial E_{D,z} = - e^2d^2t^2_\perp/\pi v^2 \sqrt{t^2_\perp+N^2_z}$. As mentioned, even in the limit $N_z\rightarrow 0$ the polarizability remains finite and reduces to $\partial P_{D,z}/ \partial E_{D,z} = - e^2d^2|t_\perp|/\pi v^2$.

\mbox{}

\subsection{Orbital magnetization response to perpendicular electric field}
\label{ssec:bilayer-AFM-M}

Next, we consider the orbital magnetization response to the perpendicular electric field. This is an example of a (linear) magnetoelectric response generally allowed in a $\mathcal I \mathcal T$-symmetric antiferromagnet, but forbidden in nonmagnetic systems which preserve $\mathcal T$. In the case of the 2D bilayer, straightforward symmetry arguments suggest that antiferromagnetic N\'eel order oriented along the perpendicular $\hat z$ direction leads to an orbital magnetization $M_z$ when an electric displacement field $E_{D,z}$ is applied. Other than the assumption of an out-of-plane N\'eel vector we will not make any particular assumptions about the form of the functions that parametrize $H_0$ in Eq.~\eqref{eq:H_k-2D}. 

We seek an expression for the magnetoelectric polarizability $\partial M_{z}/ \partial E_{D,z} $, which is obtained by applying the general result of Eq.~\eqref{eq:dM_a-4-band} to the antiferromagnetic bilayer model given by Eqs.~\eqref{eq:H_k-2D} and \eqref{eq:E_Dz}. As mentioned above, the perturbation can be rewritten as $W = w^{12}\Sigma_{12} + w^{21}\Sigma_{21} $ with $w^{12}=-w^{21} = ed/2$, while the Hamiltonian is parametrized as $\bh = (f_x, f_y, n_x, n_y,n_z)$. Substituting this into Eq.~\eqref{eq:dM_a-4-band} then yields
\begin{widetext}
\be
\frac{\partial M_{z}}{\partial E_{D,z}} = - \frac{e^2d }{\hbar}\int \frac{d^2\bk}{(2\pi)^2} \frac{\bn_\bk \cdot \partial_x\bn_\bk \times \partial_y\bn_\bk }{(f^2_{\bk}+\bn_\bk^2)^{3/2}} + 
  \frac{e^2d }{\hbar}\int \frac{d^2\bk}{(2\pi)^2} \frac{f_x(\partial_y\varepsilon_{0} \partial_x f_y  -\partial_x\varepsilon_{0}\partial_y f_y)  +  (x\leftrightarrow y) }{(f^2_{\bk}+\bn_\bk^2)^{3/2}}, \label{eq:orbME-2D}
\ee
\end{widetext}
where the first term corresponds to the geometric term and these second to the interband term. The first term was recently obtained and discussed in Ref.~\onlinecite{Radhakrishnan:arXiv2025}. The additional interband term arises here because Eq.~\eqref{eq:H_k-2D} includes a dispersion $\varepsilon_{0,\bk}$ [see Eq.~\eqref{eq:H_k-2D}], which violates the reflection condition discussed in Sec.~\ref{ssec:orb-mag}. 

The form of the geometric term suggests that even in this four-band model, for which the Berry curvature of the occupied subspace identically vanishes, the geometric term still encodes a geometric property, in this case of the vector $\bn_\bk$. In particular the form of the numerator makes this clear. As pointed out in Ref.~\onlinecite{Radhakrishnan:arXiv2025}, the integral would yield a quantized value if $f_\bk \equiv 0 $, since in this case it would express the winding of $\bn_\bk$. 

Equation \eqref{eq:orbME-2D} describes the orbital magnetization response to a perpendicular displacement field. One might wonder whether there is also a spin magnetization response, since symmetry in principle allows such response. This question is readily addressed using Eqs.~\eqref{eq:<dS_a>} and \eqref{eq:dAdL-4band}. The former defines the spin magnetization response to an electric that couples to layer pseudospin, and the latter gives the application to four-band models. It follows directly from Eq.~\eqref{eq:dAdL-4band} that the spin magnetization response vanishes, since $\bw$ and $\bm{x}$ are both zero, and only $w^{12}= -w^{21}$ and $x^{34}=x^{43}$ are nonzero. This is due to the fact that $S_z = \hbar \sigma^z/2$, which can be written as $(\hbar/2)[\tau^z\sigma^x,\tau^z\sigma^y]/2i$.

\mbox{}

\section{Polarizability of gapped Dirac fermions in 2D}
\label{sec:2D-Dirac}

This section explores a further application of the linear polarizability formulas and is specifically focused on a special class of insulators in two dimensions: insulators which are described by massive Dirac theories. In such insulators the essential electronic structure at low energies is accurately captured by a continuum Dirac Hamiltonian with a mass term, and key properties of the insulating state can be determined from this effective linearized description. The canonical example of a condensed matter system described by a Dirac Hamiltonian is graphene~\cite{CastroNeto:2009p109}, which acquires an energy gap in the presence of a sublattice potential. The latter enters the Dirac theory as a mass term. Further realizations of Dirac systems have been proposed or identified~\cite{Vafek:2014p83}, of which the surface states of topological insulators are a prominent example~\cite{Hasan:2010p3045,Qi:2011p1057}. 

Our goal here is to demonstrate---in a general setting, without reference to a specific model or material realization---that both the electric Berry phase polarizability and the orbital magnetic polarizability of insulators in this class carry signatures of the topology of 2D Dirac fermions. This is achieved in a straightforward manner by making use of the polarizability formulas obtained for two-band models. 

\begin{figure}
	\includegraphics[width=0.95\columnwidth]{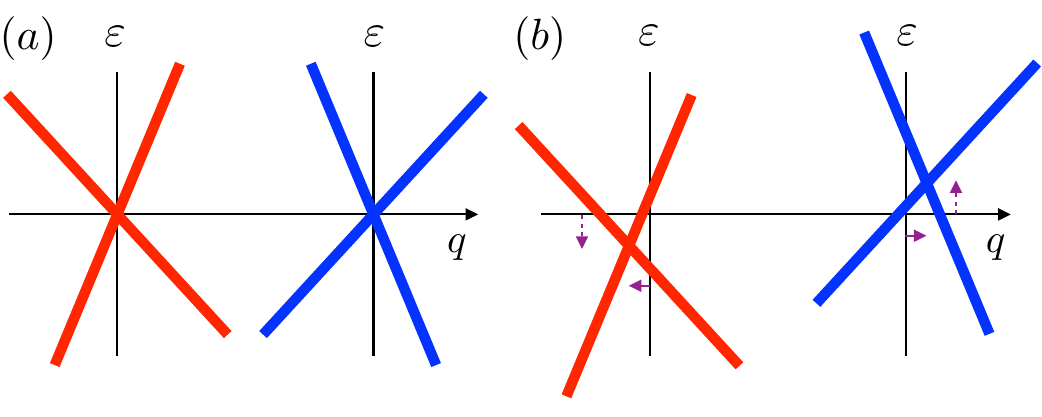}
	\caption{(a) Schematic representation of two linear Dirac band crossings described by Eq.~\eqref{eq:H_0-Dirac}. The two Dirac nodes have opposite helicity (indicated as red and blue) and can be tilted when they occur at points of low symmetry in the Brillouin zone.  (b) In the presence of a perturbation the Dirac points can be shifted in momentum and in energy, as described by Eq.~\eqref{eq:W-Dirac}.  }
	\label{fig:Dirac}
\end{figure}

The first step is to introduce the appropriate model for the unperturbed Hamiltonian. A consistent theory free of anomalies requires two flavors of fermions (i.e., ``fermion doubling'')~\cite{Semenoff:1984p2449}, which here we label $\pm$, and we introduce the unperturbed Hamiltonian for each flavor as
\be
H^\pm_0 = \pm \hbar \bv_0 \cdot \bq \pm \hbar v_D (q_x \sigma^x+q_y \sigma^y) \pm m \sigma^z, \label{eq:H_0-Dirac}
\ee
where $\hbar \bq = (\hbar q_x, \hbar q_y)$ is momentum. In each flavor sector the Hamiltonian describes a tilted Dirac cone with Dirac velocity $v_D$ and mass $m$, as shown in Fig.~\ref{fig:Dirac}(a). The tilt is caused by the first term with uniform velocity $\bv_0$, which we assume satisfies $|\bv_0| < v_D$. Symmetries may require $\bv_0$ to vanish, as is the case in graphene for instance, but here we consider the most general case and allow for a tilt. We will show below that a tilt can have an effect on the polarizability. 

A few additional remarks concerning the unperturbed Dirac Hamiltonian are useful. First, note that the masses are chosen such that insulating state has vanishing Chern number and is thus topologically trivial. Our interest here is specifically in insulators which are not Chern insulators. Second, note that the Hamiltonian does not include terms which couple the two flavors, such that it remains diagonal in flavor space and each flavor sector can be analyzed separately. The tacit assumption is that symmetries forbid a coupling between the flavors. Third, we remain agnostic about the interpretation of the Pauli matrices $\sigma^a$, which may refer to a general pseudospin degree of freedom or physical spin. The precise realization of Eq.~\eqref{eq:H_0-Dirac} is not crucial for what follows.

Next, consider the form of the perturbation. The perturbation in each flavor sector is denoted $W^\pm$ and takes the general form
\be
W^\pm  = \pm w_0 + w^x \sigma^x+w^y \sigma^y, \label{eq:W-Dirac}
\ee
where $w_0$ and $\bw = (w^x, w^y)$ are constants with dimension energy. Note that the sign of $w_0$ is opposite for the two flavors of Dirac fermions, whereas $\bw$ has the same sign. Note further that we do not include a term $w^z \tau^z$ in \eqref{eq:W-Dirac}, since a constant term proportional to $\tau^z$ is already included in the unperturbed Hamiltonian \eqref{eq:H_0-Dirac} (as a mass term), and therefore cannot constitute a symmetry breaking perturbation. The qualitative effect of both $w_0$ and $\bw $ is illustrated in Fig.~\ref{fig:Dirac}(b). As is straightforward to infer from Eqs.~\eqref{eq:H_0-Dirac} and \eqref{eq:W-Dirac}, the scalar $w_0$ shifts the two Dirac points in energy and a nonzero vector $\bw$ shifts the Dirac points in momentum. 

The next step is then to determine the electric Berry phase polarizability and the orbital magnetic polarizability for the above Dirac fermion model. The general expressions for these two polarizabilities are given by Eqs.~\eqref{eq:dP-2} and~\eqref{eq:dM_a-final}, respectively, but since the Hamiltonian for each flavor represents a two-band model, we can make straightforward use of the formulas given by Eqs.~\eqref{eq:dPdL-2-band} and~\eqref{eq:dM_a-2-band}.

\subsection{Electric (Berry phase) polarizability}

Consider first the electric Berry phase polarizability given by Eq.~\eqref{eq:dPdL-2-band}. We apply this formula to each of the two flavors separately and obtain
\begin{align}
\frac{\partial P^\pm_a}{\partial \lambda} & = e \int \frac{d^2\bk}{(2\pi)^2} \frac{m \hbar v_D \epsilon_{ab} w^b}{2(m^2 + \hbar^2v^2_Dq^2)^{3/2}}, \\
& =\frac{1}{2} \frac{e }{2\pi \hbar v_D} \epsilon_{ab} w^b \, \text{sgn}(m),
\end{align}
which shows that each flavor contributes equally. Summing these two contributions to obtain the electric polarizability of the full system yields
\be
\frac{\partial P_a}{\partial \lambda}  = \frac{e }{2\pi \hbar v_D} \epsilon_{ab} w^b \, \text{sgn}(m). \label{eq:Pa-Dirac}
\ee
It follows that the polarizability is directly proportional to $\bw$, i.e., the vector which corresponds to the shift of the Dirac points in momentum space, and furthermore only depends on the sign of the mass $m$. The latter is a signature of the topological nature of the response described by Eq.~\eqref{eq:Pa-Dirac}. In fact, the polarizability of Eq.~\eqref{eq:Pa-Dirac} is fully consistent with---and essentially a restatement of---the electromagnetic response theory of topological semimetals in 2D, as derived by Ramamurthy and Hughes (RH)~\cite{Ramamurthy:2015p085105} (later rederived in Ref.~\onlinecite{Yoshida:2023p075160}). In Ref.~\onlinecite{Ramamurthy:2015p085105} RH showed that a Dirac semimetal in 2D has an electric polarization that is proportional and perpendicular to the Dirac dipole moment, i.e., the separation of the Dirac points in momentum space. Within the Dirac model considered here this vector is given by $\lambda \bw / \hbar v_D$, which leads to the polarizability given by \eqref{eq:Pa-Dirac}. Note that the polarizability of Eq.~\eqref{eq:Pa-Dirac}---the change in polarization---does not suffer from subtleties and ambiguities that plague the determination of the polarization itself~\cite{Ramamurthy:2015p085105}. 

A classic example of the polarization response described by Eq.~\eqref{eq:Pa-Dirac} is graphene in the presence of strain. Strain is well-known to couple to low-energy Dirac electrons of graphene as a pseudo-gauge field, thus shifting the position of the Dirac nodes in momentum space~\cite{Morpurgo:2006p196804,Guinea:2008p075422,CastroNeto:2009p109,Guinea:2010p30}. In this instance $\lambda$ should therefore be interpreted as a strain perturbation and $m$ is an inversion symmetry breaking mass coming from a sublattice potential~\cite{CastroNeto:2009p109}. 

Another manifestation of Eq.~\eqref{eq:Pa-Dirac} was recently pointed out in the context of symmetry-enforced 2D Dirac semimetals which develop $\mathcal I \mathcal T$-symmetric antiferromagnetic order~\cite{Young:2015p126803,Wang:2017p115138,Smejkal:2017p106402,Venderbos:arXiv2025}. Dirac semimetals of this kind are characterized by fourfold degenerate linear crossings at high symmetry points of the Brillouin zone (i.e., the two nodes depicted in Fig.~\ref{fig:Dirac} are degenerate, with no tilt), which acquire a gap in the presence of antiferromagnetic order. A Berry polarization is then induced by the application of a Zeeman field, such that in this instance Eq.~\eqref{eq:Pa-Dirac} describes a magnetoelectric polarizability~\cite{Venderbos:arXiv2025}. 

These two examples show that both the nature of the Dirac points and the nature of the perturbation may differ from system to system. In graphene the Dirac points do not rely on spin at all, whereas in the symmetry-enforced Dirac semimetals spin-orbit coupling plays a key role in realizing Dirac points. Similarly (and correspondingly), strain breaks neither time-reversal nor inversion symmetry, whereas a Zeeman field clearly breaks time-reversal symmetry and couples to spin. The Dirac theory description introduced here, and the polarizability derived from it, are sufficiently general to apply to all cases. 

\subsection{Orbital magnetic polarizability}

Next, consider the orbital magnetic polarizability. For each of the two flavors of Dirac fermions the polarizability can be calculated using Eq.~\eqref{eq:dM_a-2-band}, which yields
\be
\frac{\partial M^\pm_z}{\partial \lambda} = - \frac{1}{2 }\frac{e}{2\pi \hbar }\left( w_0 \mp \frac{\bv_0 \cdot \bw}{v_D} \right)\text{sgn}(m).  \label{eq:Mz-Dirac-pm}
\ee
The first term in parenthesis on the right hand side ($w_0$) is the geometric contribution and the second term ($\bv_0 \cdot \bw/v_D$) is the interband contribution. We thus find that the geometric contribution is of the same sign, whereas the interband contribution is of opposite sign. As a result, while a single flavor of Dirac fermions has both a geometric and an interband contribution, the full orbital magnetic polarizability of this Dirac system is given by
\be
\frac{\partial M_z}{\partial \lambda} = - w_0 \frac{e}{2\pi \hbar }  \text{sgn}(m), \label{eq:Mz-Dirac}
\ee
and is thus entirely due to the geometric term.

The polarizability given by \eqref{eq:Mz-Dirac} expresses the fact an orbital magnetization is induced when a perturbation shifts the Dirac fermions in energy. Such a response was first described by RH in Ref.~\onlinecite{Ramamurthy:2015p085105}. We obtain our result via a different route: we have first determined the general orbital magnetization response to symmetry breaking perturbations and then applied the derived formulas to the case of Dirac insulators. In this way, by focusing on the orbital magnetic \emph{polarizability} (rather than the orbital magnetization directly), we find that for a single Dirac fermion there are in fact two contributions to the polarizability, as seen from Eq.~\eqref{eq:Mz-Dirac-pm}. The first contribution comes from the geometric term and is determined by Berry curvature and the energy shift of the Dirac point. The second contribution comes from the interband term, which, in the case of two-band models, reduces to the Berry curvature polarizability. The latter is proportional to $\bw$, the shift in momentum. This second contribution further requires that the Dirac points are tilted, as expressed by $\bv_0$. This second contribution has not been discussed before. For the elementary Dirac theory discussed here, defined by two Dirac points of opposite helicity, the total contribution from the Berry curvature polarizability vanishes, but one may imagine more general theories with multiple sets of Dirac points for which the net contributions do not vanish.

\section{Hall vector polarizability in a 2D altermagnet}
\label{sec:apply-altermagnet}

\begin{figure}
	\includegraphics[width=\columnwidth]{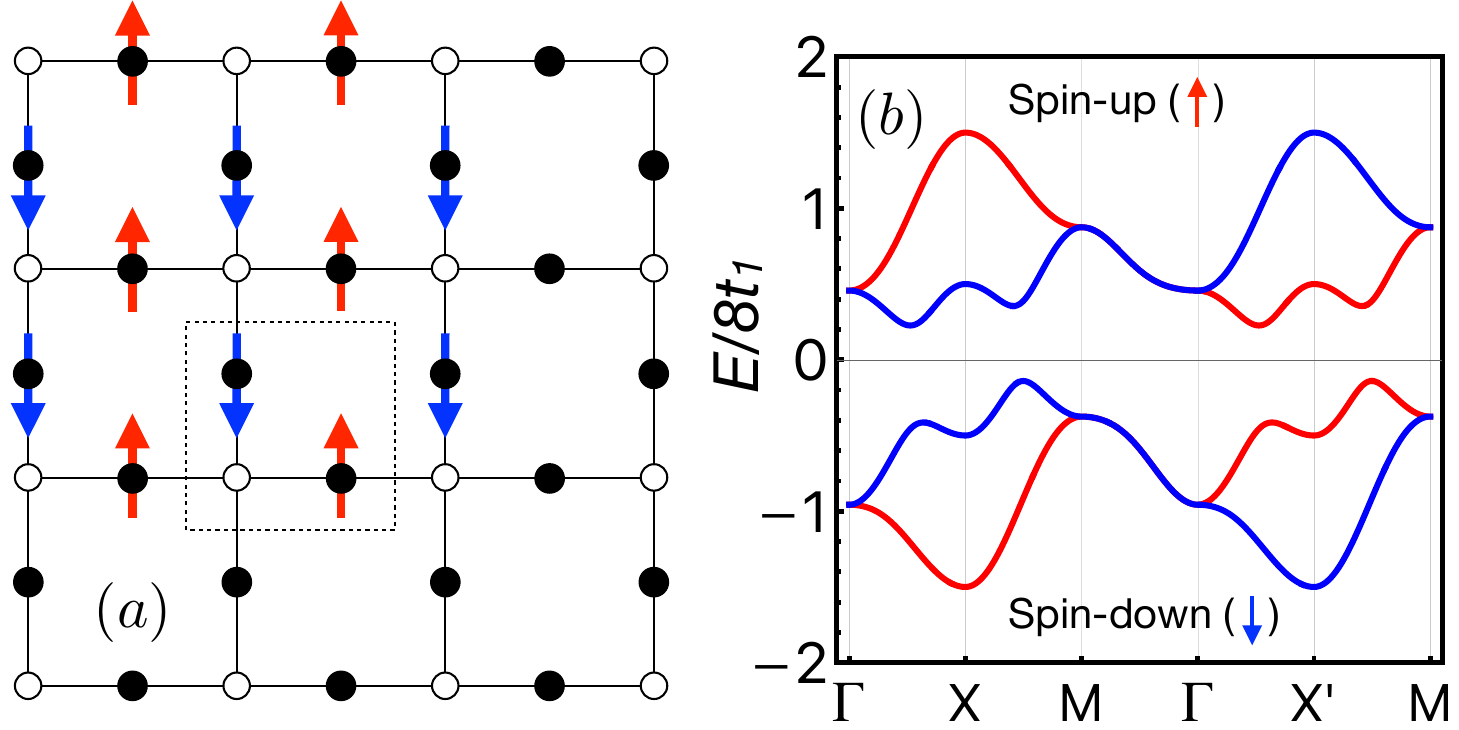}
	\caption{(a) The unit cell of the two-dimensional Lieb lattice has two magnetic sites, shown in black and labeled $A$ and $B$, and one non-magnetic site, shown in white. The ordered moments of the altermagnetic  collinear N\'eel state are indicated as red and blue arrows. (b) Energy spectrum of the Lieb lattice model defined by Eq.~\eqref{eq:Lieb-def}. We have used the parameters $(t_0,t_d,t_{\text{SO}},N_z) = (0.5t_1,2.0t_1,0.75t_1,4.0t_1)$; blue and red bands correspond to $\sigma=\up$ and $\sigma=\down$, respectively. }
	\label{fig:Lieb}
\end{figure}

In this section we consider a final application of the polarizability formulas. We focus specifically on the Hall vector polarizability introduced in Sec.~\ref{sec:Hall} and examine a simple lattice model in 2D for which the Hall vector polarizability is of interest. The model in question describes a Lieb lattice magnet, shown in Fig.~\ref{fig:Lieb}, which has become a paradigmatic minimal model~\cite{Brekke:2023p224421,McClarty:2024p176702,Schiff:2025p109,Antonenko:2025p096703,Roig:2024p144412,Yershov:2024p144421,Durrnagel:2025p036502} for a type of magnetism known as altermagnetism~\cite{Smejkal:2022p031042,Smejkal:2022p040501}. One of the reasons altermagnets have attracted attention is their response to applied strain, for instance their magnetization response to strain (i.e., piezomagnetism)~\cite{Ma:2021p2846,Aoyama:2024pL041402,McClarty:2024p176702,Takahashi:2025p184408,Belashchenko:2025p086701,Naka:2025p083702,Khodas:arXiv2025}, which has placed emphasis on the potential for strain engineering the properties of altermagnets~\cite{Khodas:arXiv2025,Zhai:2025p174411,Karetta:2025p094454,Fu:arXiv2025}. Motivated by this effort, in this section we study how the Berry curvature---or, equivalently, the Hall vector---of a 2D altermagnet responds to the application of strain, using the notion of Hall vector polarizability.

As shown in Fig.~\ref{fig:Lieb}, the Lieb lattice is a square lattice with a three-site unit cell. Two sites correspond to magnetic sites (shown in black) and one is a nonmagnetic site (shown in white). The presence of a nonmagnetic site is responsible for the anisotropic crystallographic environment seen by the magnetic sites, and hence for altermagnetism. While a full microscopic model may include all three sites~\cite{Brekke:2023p224421}, here we construct a model for the magnetic sites only, which are labeled $A$ and $B$~\cite{Antonenko:2025p096703}. 

The Lieb lattice altermagnet is defined by compensated collinear magnetic order with anti-aligned moments on the $A$ and $B$ sublattice (see Fig.~\ref{fig:Lieb}). Since neither translations nor spatial inversion exchange the (magnetic) sublattices, the energy bands are allowed to be spin-split in a nonuniform way, which is a key characteristic of altermagnetism. We include the effect of spin-orbit coupling and assume easy-axis anisotropy with ordered moments aligned along the $\hat z$ direction. With the N\'eel vector oriented along the $\hat z$ direction an in-plane mirror symmetry $\mathcal M_z  :  z\rightarrow -z$ and a combined $\mathcal  T \mathcal C_{4z}$ symmetry are preserved ($ \mathcal  T$ denotes time-reversal), which forbids an overall magnetization and thus guarantees full compensation of the ordered moments. 

A microscopic electronic model is constructed by introducing the electron destruction operators $c_{\bk \alpha \sigma}$, where, as in Sec.~\ref{sec:apply-1D}, $\alpha=A,B$ labels the two sublattices and $\sigma=\up,\down$ denotes spin. The full Hamiltonian is defined as $H = \sum_{\bk} c^\dagger_\bk \mathcal H_\bk c_\bk$, with $\mathcal H_\bk = H_{0,\bk} + \lambda W_\bk$. As a result of the mirror symmetry $\mathcal M_z$ the Hamiltonian commutes with the mirror operator at each $\bk$, and this implies that the Hamiltonian is diagonal in spin space; the spin species are not allowed to couple. The Hamiltonian can be examined in each spin sector separately and thus reduces to two decoupled two-band models. Hence, the unperturbed Hamiltonian has the general structure
\be
H^\sigma_{0}  = h_0 +\bh^\sigma  \cdot \btau, \label{eq:H_0^sigma}
\ee
where $\sigma$ labels the spin sectors and $\tau^z = \pm 1$ corresponds to the sublattices $A$ and $B$. We choose the functions $h_0$ and $\bh^\sigma$ as $h_0 = -2t_0 (\cos k_x + \cos k_y)$ and 
\be
\begin{split}
h^{x,\sigma} & = -4t_1 \cos(k_x/2) \cos(k_y/2),\\
h^{y,\sigma} & = 4\sigma t_{\text{SO}} \sin (k_x/2) \sin (k_y/2), \\
h^{z,\sigma} & = -2t_d (\cos k_x - \cos k_y) + \sigma N_z.
\end{split} \label{eq:Lieb-def}
\ee
(See Refs.~\onlinecite{Brekke:2023p224421,Antonenko:2025p096703}.) Here $t_1$ is a nearest neighbor hopping between the sublattices and $t_0$ is a uniform second nearest neighbor hopping within the sublattices; $t_d$ is an anisotropic intra-sublattice hopping reflecting the different local crystal environment seen by the sublattice. We have further included a spin-orbit coupling term parametrized by $t_{\text{SO}} $, and $N_z$ is the magnetic N\'eel order parameter.

\begin{figure}
	\includegraphics[width=\columnwidth]{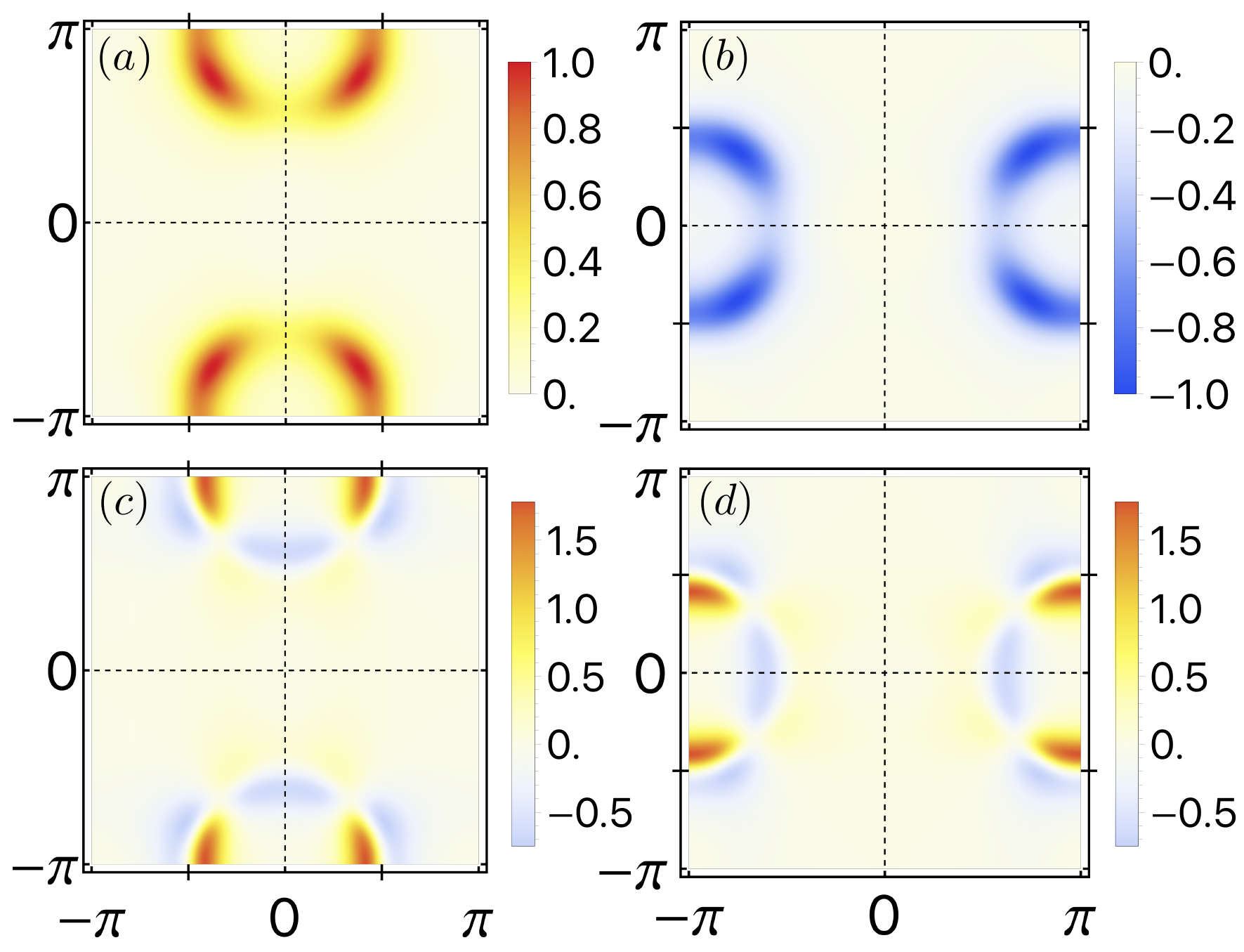}
	\caption{Berry curvature and Berry curvature polarizability of the Lieb lattice model. (a) and (b) show the Berry curvature $\Omega_{xy}$ of the valence band in the $\sigma=\up$ and $\sigma=\down$ sector, respectively. (c) and (d) show the Berry curvature (or Hall vector) polarizability $\partial_\lambda \Omega_{xy}$ of the valence band in the $\sigma=\up$ and $\sigma=\down$ sector, respectively. The Hall vector polarizability is calculated using the formula of Eq.~\eqref{eq:dOmega-2band}, with model parameters set to $(t_0,t_d,t_{\text{SO}},N_z) = (0.5t_1,2.0t_1,0.75t_1,4.0t_1)$ [See Eq.~\eqref{eq:Lieb-def}].  }
	\label{fig:Lieb-Berry}
\end{figure}

Consider next the strain perturbation. We follow Refs.~\onlinecite{Takahashi:2025p184408} in modeling the effect of strain. The strain perturbation takes the general form
\be
W^\sigma = w_0+  w^{z} \tau^z , \label{eq:W_k}
\ee
and is equal in each spin sector. Here $w_0$ is a uniform strain-induced anisotropic hopping and $w^{z} $ is a hopping anisotropy with opposite sign for the two sublattices. These functions of $\bk$ are chosen as
\be
\begin{split}
w_0 & = -2t_0 (\cos k_x - \cos k_y), \\
 w^{z} &= -2t_d(\cos k_x + \cos k_y),
\end{split} \label{eq:Lieb-strain}
\ee
such that the intra-sublattice second nearest neighbor hopping amplitude along the $\hat x$ direction becomes $(t_0 \pm t_d)(1+\lambda)$ in the presence of strain, whereas the hopping amplitude along the $\hat y$ direction becomes $(t_0 \mp t_d)(1-\lambda)$.

The energy bands of the Lieb lattice altermagnet in the absence of strain are shown in Fig.~\ref{fig:Lieb} for a representative set of parameters. Our focus will be on the two valence bands in each spin sector, which are color coded red and blue for spin-up and -down, respectively. Consider first the Berry curvature of the unstrained model. The Berry curvature distribution of the two valence bands is shown in Figs.~\ref{fig:Lieb-Berry}(a) and~\ref{fig:Lieb-Berry}(b). Panels (a) and (b) correspond to $\sigma=\up$ and $\sigma=\down$, respectively. The two valence bands are related by $\mathcal T \mathcal C_{4z}$ symmetry, i.e., the product of time-reversal and fourfold rotation, and the Berry curvature clearly reflects this. A notable feature is the concentration of Berry curvature in the vicinity of points located on the $k_x=\pi$ and $k_y=\pi$ lines. As was pointed out in Ref.~\onlinecite{Antonenko:2025p096703}, in the limit of vanishing spin-orbit coupling ($t_{\text{SO}} =0$) this Lieb lattice model has symmetry-protected Dirac band crossings on the Brillouin zone boundary (as long as $|N_z / 4t_d|<1$), which acquire a gap when $t_{\text{SO}} \neq 0$. The location of these Dirac points is indicated on the $k_y=\pi$ and $k_x=\pi$ lines in Figs.~\ref{fig:Lieb-Berry}(a) and~\ref{fig:Lieb-Berry}(b), respectively. The distribution of Berry curvature can be understood from these Dirac points. Note that here we have chosen a large gap by setting $t_{\text{SO}} =0.75t_1$, which explains the smoothening and shape of the Berry curvature seen in panels (a) and (b) of Fig.~\ref{fig:Lieb-Berry}.

\begin{figure}
	\includegraphics[width=\columnwidth]{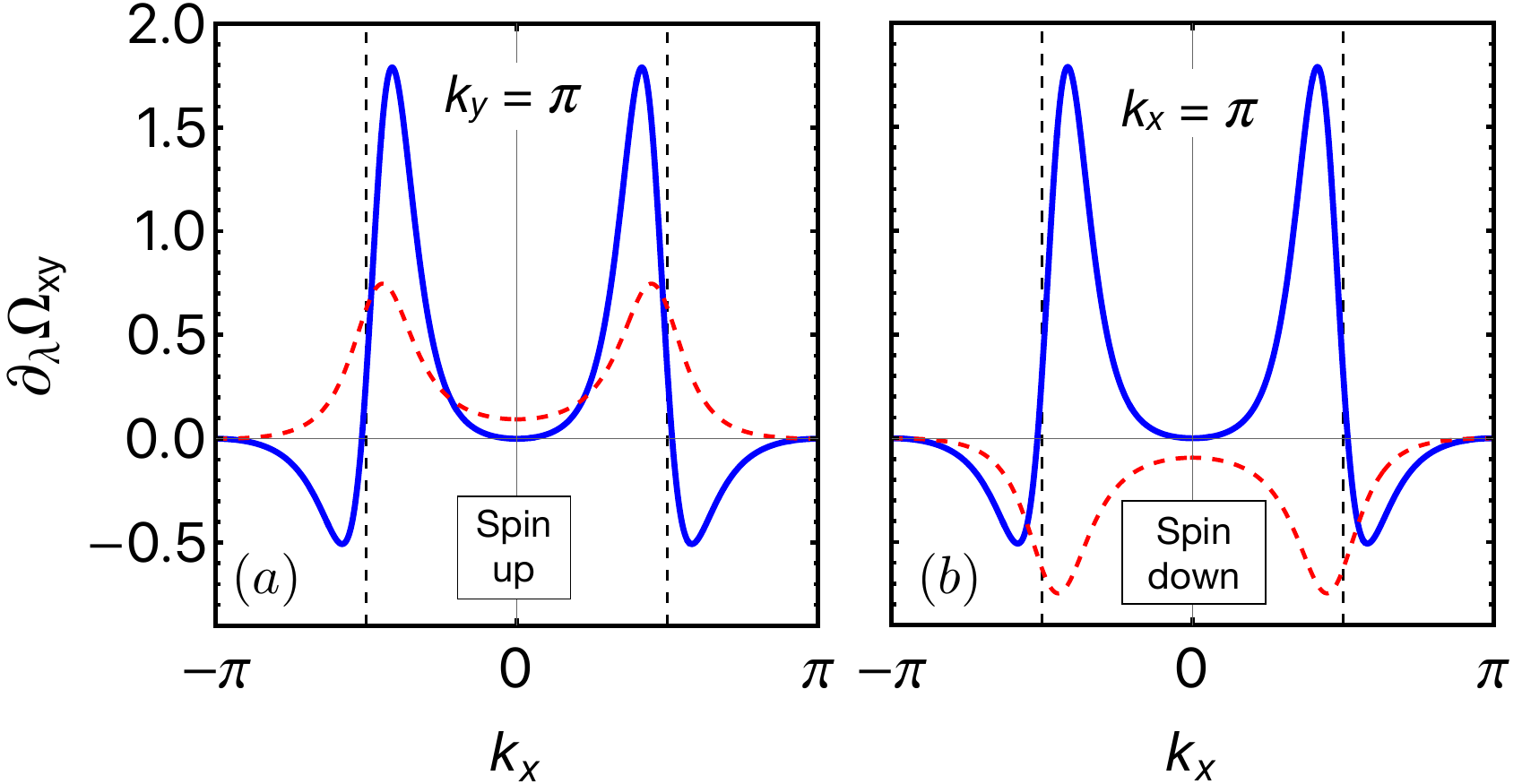}
	\caption{(a) Berry curvature polarizability $\partial_\lambda \Omega_{xy}$ on the $k_y=\pi$ line of the Lieb lattice altermagnet $\sigma=\up $ valence band (shown in blue). The dashed red curve shows the Berry curvature $\Omega_{xy}$. The lattice constant has been set to unity. (b) Same as in (a) for the $\sigma=\down $ valence band on the $k_x=\pi$ line. In both panels we have used the same model parameters as in Fig.~\ref{fig:Lieb-Berry}. }
	\label{fig:Lieb-Berry-cut}
\end{figure}

We then examine the Berry curvature polarizability of the two valence bands. To compute the polarizability in each spin sector, we make use of the general formula for two-band models given by Eq.~\eqref{eq:dOmega-2band}. In this specific case $\bh^\sigma$ is given by \eqref{eq:Lieb-def} and $\bw = w^z \hat\bz$, with $w^z$ given by \eqref{eq:Lieb-strain}. The Berry curvature polarizability $\partial_\lambda \Omega_{xy}$ of the spin-up and spin-down bands is shown in Figs.~\ref{fig:Lieb-Berry}(c) and~\ref{fig:Lieb-Berry}(d), respectively. Two key features are important to highlight. First, whereas the Berry curvature of one valence band can be obtained from the other by fourfold rotation and sign inversion, as required by $\mathcal T \mathcal C_{4z}$ symmetry, the Berry curvature polarizabilities are related by fourfold rotation only. The polarizability, i.e., the change of Berry curvature in response to strain, thus has the same sign. This reflects the fact that strain breaks the $\mathcal T \mathcal C_{4z}$ symmetry and allows an imbalance of Berry curvature between the two bands. This imbalance is the reason why the Berry curvature polarizability is directly related to the elasto-Hall conductivity~\cite{Takahashi:2025p184408}.

The second feature can be observed in the vicinity of the (gapped) Dirac points, where the Berry curvature is concentrated. Whereas the Berry curvature has the same sign in this region (i.e., positive and negative for spin-up and spin-down, respectively), the polarizability has the structure of a dipole, with a sign change on a nodal line intersecting the Dirac points. To show this more clearly, we present line cuts of the Berry curvature polarizability in Fig.~\ref{fig:Lieb-Berry-cut}. The Berry curvature polarizability of the spin-up band on the $k_y=\pi$ line is shown in Fig.~\ref{fig:Lieb-Berry-cut}(a); a similar plot is shown in panel (b) for the spin-down band. The red dashed curves show the Berry curvature itself, which is of opposite sign for the two bands. The location of the Dirac points, as determined from the $t_{\text{SO}} = 0$ limit, is indicated by vertical dashed lines. Figure~\ref{fig:Lieb-Berry-cut} suggests that the Berry curvature can be viewed as a ``monopole'' centered on the Dirac point, whereas the polarizability can be viewed as ``dipole''. This is consistent with the general result obtained in Sec.~\ref{sec:Hall} that the Berry curvature (or Hall vector) polarizability is equal to a generalized curvature dipole $\sim \partial_b \Omega_{c\lambda}$, see Eqs.~\eqref{eq:Berry-sum} and \eqref{eq:dOmega_ab}. Hence, both Fig.~\ref{fig:Lieb-Berry} and Fig.~\ref{fig:Lieb-Berry-cut} clearly demonstrate the nature of the Berry curvature polarizability as a curvature dipole.

\mbox{}

\section{Discussion and Conclusion \label{sec:discuss}}

In this paper we have revisited the polarization and magnetization responses of crystalline electronic insulators. Our primary concern has been the Berry phase polarization and orbital magnetization responses---the two core concepts of the ``modern theory''---and our main result is a band theory-based formulation of these responses to arbitrary (lattice-periodic) perturbations. This extends previous efforts to describe response properties of solids by offering a generalized and more broadly applicable formalism for studying the effect of external perturbations. Our formulation furthermore exposes a number of interesting connections between different types of responses, such as the relationship between the generalized curvature describing Berry phase polarization responses and the Berry curvature or Hall vector polarizability. The latter is also shown to enter the expression for the orbital magnetization response when certain conditions are met. 

The approach we have chosen to study responses is particularly suited for the analysis of lattice tight-binding Hamiltonians, although it is not limited to this class of models. Tight-binding models generally provide key insight into the essential electronic properties of real materials, and it is therefore useful to expand the array of analysis tools for such models. With the application to tight-binding models in mind, we have derived simplified polarizability formulas for special classes of such models: generic two-band models and four-band models with a symmetry requirement. In many instances the models used to examine specific systems fall in one of these two classes. Examples have been discussed in Secs.~\ref{sec:apply-1D} and \ref{sec:2D-bilayer-AFM}, where we have considered $\mathcal I \mathcal T$-symmetric antiferromagnets in one and two dimensions, respectively. A variety of different polarizabilities of the kind considered in this work can be computed for these models, showcasing the broad applicability and utility of the polarizability formulas.

We have also studied responses of generalized pseudospin observables to external electric and magnetic fields. In a sense these can be viewed as dual to the Berry phase polarization and orbital magnetization responses. Whereas in the latter case we have been interested in determining the ``orbital motion'' response of electrons to external perturbations that couple to spin or pseudospin degrees of freedom, for instance (i.e., lattice-periodic perturbations that do not directly couple to orbital motion), the former refers to the spin or pseudospin responses resulting from perturbations that couple directly to the orbital motion of electrons (i.e., electric and magnetic fields). This duality is reflected in the Maxwell relations that can be established for such responses. In Sec.~\ref{sec:apply-ME} we have done so explicitly in the context of magnetoelectric polarizabilities, but it is clear from Eqs.~\eqref{eq:dA_dBa} and~\eqref{eq:dM_a-final} that this may be achieved more generally in a straightforward way.

A key result of this paper is the detailed analysis of the Hall vector polarizability, i.e., the response of the Berry curvature---a band geometric object---to external perturbations. We have in particular shown that the Hall vector polarizability can be understood as a generalized curvature dipole, where the generalized curvature is the same curvature that describes the Berry phase polarization response to the perturbation in question. This interpretation is likely to be useful in a number of practical settings, since the Berry curvature plays a role in various known responses, in particular in the anomalous Hall response. The Hall vector polarizability is then of interest when symmetries forbid a response, but breaking these symmetries by applying external perturbations may unlock the nontrivial effects of Berry curvature. The further examination of such scenarios is left for future work. 


\section*{Acknowledgements}

I am greatly indebted to Carmine Ortix, Paola Gentile, Harini Radhakrishnan and Beryl Bell for collaborations on projects related to this paper, as well as for stimulating conversations that have motivated and influenced it. I am particularly grateful to Harini Radhakrishnan and Beryl Bell for a careful reading of the manuscript. I also gratefully acknowledge insightful discussions with Rafael Fernandes, Daniel Agterberg, and Taylor Hughes on matters directly or indirectly related to this work. This research was supported by the U.S. Department of Energy under Award No. DE-SC0025632.

\appendix

\section{Four-band model with symmetry \label{app:4-band}}

In the main text we examine a special class of four-band Hamiltonians, namely those with $\mathcal I \mathcal T $ symmetry. The purpose of this Appendix is to provide further details on the analysis of this class of Hamiltonians. 

In general, an $N$-band Hamiltonian can be expanded in $N^2-1$ Hermitian generators with real coefficients (up to a term proportional to the identity). In the case of $N=4$ (i.e.,  $4\times 4$ Hamiltonians) there are $15$ such Hermitian matrices. As mentioned in the main text, with the constraint of $PT$ symmetry a general four-band Hamiltonian can be expanded using five mutually anti-commuting Hermitian generators, which we denote $\Gamma_\alpha$. These satisfy $\{\Gamma_\alpha, \Gamma_\beta \} = 2\delta_{\alpha\beta}\mathbb{1}$. The Hamiltonian can be expressed as $H = h^\alpha \Gamma_\alpha= \bh \cdot \bGamma$, where $\bh$ is a five-component real (and momentum-dependent) vector. The spectrum has two branches $\pm \varepsilon$, with $\varepsilon = |\bh| = \sqrt{\bh\cdot \bh}$, and each branch is manifestly twofold degenerate---as required by $PT$ symmetry.

The form of the Hamiltonian gives rise to a simple expression for the projectors onto the two bands. Note that in this work we tacitly assume the two branches are separated by an energy gap, meaning that $\varepsilon =\varepsilon_\bk \neq 0$ for any momentum $\bk$. The lower band with energy $- \varepsilon$ is then the occupied band. The projector onto the occupied band is 
\be
P =  \frac12(\mathbb{1}- \bh \cdot \bGamma/\varepsilon )=  \frac12(\mathbb{1}- H/\varepsilon ). \label{app:project-4band}
\ee
which directly follows from $H^2 = \varepsilon^2 \mathbb{1}$.

In addition to the Hamiltonian, we also require an expression for the perturbation $W$ and the observable $X$. These must be described by Hermitian matrices, but in general these are not subject to any symmetry constraints. Therefore, in general these cannot be expanded in the five $\Gamma$-matrices only. Instead, the most general form of the perturbation $W$ (and similarly the observable $A$) is
\be
W  = w^\alpha \Gamma_\alpha + w^{\alpha \beta} \Sigma_{\alpha \beta}, \label{app:W-Gamma}
\ee
where $\Sigma_{\alpha \beta}$ are Hermitian matrices defined as
\be
 \Sigma_{\alpha \beta} = \frac{1}{2i} [\Gamma_\alpha,\Gamma_\beta].
\ee
Clearly, these are antisymmetric with respect to $\alpha$ and $\beta$, i.e., $ \Sigma_{\alpha \beta}=- \Sigma_{\beta\alpha }$, such that there are ten such matrices. Combined with the five $\Gamma$-matrices these span the space of $4\times 4$ Hermitian matrices, and it must therefore be possible to write $W$ (as well as $X$) in the form of \eqref{app:W-Gamma}. 

In Sec.~\ref{sec:2D-bilayer-AFM} a particular representation of the $\Gamma$-matrices is used to describe a four-band model of a bilayer antiferromagnet. In this representation the $\Gamma$-matrices are given by 
\be
(\Gamma_1,\Gamma_2,\Gamma_{3},\Gamma_{4},\Gamma_{5}) = (\tau^x,\tau^y,\tau^z\sigma^x,\tau^z\sigma^y,\tau^z\sigma^z),
\ee
where $\tau^{x,y,z}$ and $\sigma^{x,y,z}$ are Pauli matrices. The general (unperturbed) Hamiltonian of Eq.~\eqref{eq:H_k-2D} is written as an expansion in these matrices. Note that the Hamiltonian also has a uniform piece proportional to the identity, given by $\varepsilon_{0,\bk} $, which (uniformly) affects the energies but does not affect the eigenstates. In particular, the two spectral branches are given by
\be
\ve_{ 1} = \varepsilon_{0} - \ve, \quad  \ve_{2} = \varepsilon_{0} + \ve
\ee
with $\ve =( f_x^2 +f^2_y + \bn^2)^{1/2}$. The projectors onto the respective eigenspaces are given by 
\be
P_{1,2} = \frac12\left(\mathbb{1}  \mp  \frac{f_{x} \tau^x+ f_{y} \tau^y+  \tau^z \bn \cdot \bsigma}{\ve} \right),
\ee
such that $P_1$, the projector onto the occupied subspace, is indeed equal to \eqref{app:project-4band}. 

In Sec.~\ref{sec:2D-bilayer-AFM} we have considered a perturbation of the form $W= ed\tau^z$. Here we show how to connect the explicit form of $W$ to the general expansion of Eq.~\eqref{app:W-Gamma}. Note that $\tau^z$ is not one of the five $\Gamma$-matrices and therefore $w^\alpha =0$ in Eq.~\eqref{app:W-Gamma}. It is straightforward to obtain that
\be
\tau^z = \frac{1}{2i} [\tau^x,\tau^y]= \frac{1}{2i} [\Gamma_1,\Gamma_2] = \Sigma_{12} =-\Sigma_{21} , 
\ee
from which it follows that $W= w^{12} \Sigma_{12} +w^{21}\Sigma_{21} $ with $w^{12} =-w^{21} = ed/2$.


\mbox{}

\section{Berry connection polarizability} 
\label{app:berry-connection}

In this Appendix we point out that it is possible to obtain the Berry connection polarizability from Eq.~\eqref{eq:dP-3} by using the appropriate form of the perturbation $W$ when an electric field is applied. The Berry connection polarizability describes the Berry phase polarization response to an applied electric field~\cite{Gao:2014p166601,Xiao:2022p086602}. 

As discussed in Sec.~\ref{ssec:spin-mag}, an electric field enters the Hamiltonian as $H_E = e \bE \cdot \br$. This does not fit into our simple perturbative framework, since that (tacitly) assumes a lattice-periodic perturbation. Nonetheless, we can make use of Eq.~\eqref{eq:dP-3} by taking~\cite{Nunes:2000p155107} [see also Eq.~\eqref{eq:r_a^nm}]
\be
r_b^{mn} \equiv A_b^{mn} = i \langle u^0_m | \partial_b u^0_n\rangle,
\ee
where $r_b^{mn}$ are the position matrix elements and $A_b^{mn}$ is the interband Berry connection of the unperturbed bands.  

The polarizability obtained from a perturbation of this form is $\partial P_a / \partial E_b$, i.e., the Berry phase polarization response in direction $a$ to an electric field in direction $b$. As discussed in Sec.~\ref{ssec:spin-mag}, upon further rewriting $A_a^{mn} = -i \hbar v_a^{mn}/(\varepsilon_{m}-\varepsilon_n)$, with $\hbar v_a = \partial_a H_0$, one obtains from Eq.~\eqref{eq:dP-3}
\be
\frac{\partial P_a }{\partial E_b} =  -\hbar ^2  G_{ab}, \label{eq:dPdE}
\ee
with $G_{ab}$ given by
\be
G_{ab} = \int [d\bk] \sum_{n,m}  \frac{2\text{Re}[ v_a ^{nm} v_b ^{mn}]}{(\varepsilon_{n}-\varepsilon_m)^3}. \label{eq:G_ab}
\ee
The quantity $G_{ab} $ has been referred to as the  Berry connection polarizability and is directly related to the quantum metric. 
 
\section{Orbital magnetic polarizability \label{app:orb-mag}}

In this Appendix we provide details of the derivation of the orbital magnetic polarizability presented in Sec.~\ref{sec:magnetic}. Specifically, we show how Eq.~\eqref{eq:dM_a-final} is obtained from \eqref{eq:dM_a}. We also discuss an alternative derivation of Eq.~\eqref{eq:dM_a-final}, which is based on an approach taken in Ref.~\onlinecite{Malashevich:2010p053032} to compute the orbital magnetoelectric polarizability (i.e., the orbital magnetization response to an electric field).  

\subsection{Derivation of Eq.~\eqref{eq:dM_a-final}}

To derive Eq.~\eqref{eq:dM_a-final}  from \eqref{eq:dM_a}, we proceed in three steps. First, we combine the first and fourth term in the integrand of \eqref{eq:dM_a} to obtain
\begin{widetext}
\begin{multline}
\text{Im}  \text{Tr} [(\partial_b P)  Q W Q  (\partial_c P)]   + \text{Im}  \text{Tr} [W(\partial_b  P)Q (\partial_cP)]  
=\text{Im}  \text{Tr} [    (\partial_c P)(\partial_b P)QW ]   + \text{Im}  \text{Tr} [WP(\partial_b  P) (\partial_cP)] \\
= \text{Im}  \text{Tr} [W(P-Q) \partial_b  P \partial_cP].
 \label{app:dM_a}
\end{multline}
To arrive at the last line we have used the relations $ \partial_a  P = - \partial_a  Q$ and $Q\partial_a  P =  -(\partial_aQ)  P$ multiple times, relations which follow from $\mathbb{1} = P+Q$ and $PQ=0$. The final term in the last line is equal to the first term in Eq.~\eqref{eq:dM_a-final}.

The second step is to perform integration by parts on the terms which involve momentum derivatives of $\mathcal P^{(1)}$. There are two such terms, and integration by parts yields
\begin{multline}
\int[d\bk] \left(  \text{Im}  \text{Tr} [\partial_b \mathcal P^{(1)} Q H_0   Q  (\partial_c  P)]+ \text{Im}  \text{Tr} [H_0(\partial_b \mathcal P^{(1)}) Q (\partial_c P)] \right) = \\
- \int[d\bk]\left( \text{Im}  \text{Tr} [ \mathcal P^{(1)} (\partial_b Q) H_0   Q  (\partial_c  P)] + \text{Im}  \text{Tr} [ \mathcal P^{(1)} Q H_0   (\partial_b Q)   (\partial_c  P)] +\text{Im}  \text{Tr} [ H_0\mathcal P^{(1)} (\partial_b Q )(\partial_c P)]\right) \\
- \int[d\bk]  \left( \text{Im}  \text{Tr} [ \mathcal P^{(1)}  Q (\partial_bH_0  ) Q  (\partial_c  P)] + \text{Im}  \text{Tr} [(\partial_bH_0)  \mathcal P^{(1)} Q (\partial_c P)]\right) . \label{app:int-parts}
\end{multline}
On the right hand side we have ignored terms featuring double derivatives $\partial_b\partial_c  P$, since these vanish when contracted with $\epsilon_{abc}$. On the right hand side we have further collected the terms into two groups. The second group involves derivatives of the Hamiltonian $H_0$, which can be further rewritten as
\begin{multline}
\text{Im}  \text{Tr} [ \mathcal P^{(1)}  Q (\partial_bH_0  ) Q  (\partial_c  P)]  + \text{Im}  \text{Tr} [(\partial_bH_0)  \mathcal P^{(1)} Q (\partial_c P)] 
=\text{Im}  \text{Tr} [ P\mathcal P^{(1)}  Q (\partial_bH_0  )\partial_c  P]  + \text{Im}  \text{Tr} [P \mathcal P^{(1)} Q (\partial_c P)(\partial_bH_0) ]\\
=\text{Im}  \text{Tr} [ P\mathcal P^{(1)}  Q \{ \partial_b H_0 , \partial_c P\}] . \label{app:rewrite}
\end{multline}
Here we have again used $ \partial_a  P = - \partial_a  Q$ and $Q\partial_a  P =  -(\partial_aQ)  P$.  Substituting the expression for $\mathcal P^{(1)}  $ in the final line of \eqref{app:rewrite}, we arrive at
\be
\text{Im}  \text{Tr} [ P\mathcal P^{(1)}  Q \{ \partial_bH_0  ,\partial_c  P\}] \\
= -\sum_{n,m}\frac{\text{Im}  \text{Tr} [ WP_n \{ \partial_bH_0  ,\partial_c  P\}P_m]}{\varepsilon_n-\varepsilon_m},
\ee
which is the second term of Eq.~\eqref{eq:dM_a-final}. 

The third and final step is to demonstrate that all remaining terms of Eq.~\eqref{eq:dM_a} of the main text, as well as the terms in the second line of Eq.~\eqref{app:int-parts} (i.e., the additional terms produced by integration by parts), vanish. This may be achieved in a relatively straightforward way by using properties of $\mathcal P^{(1)} $, i.e., the first order correction to the projector, and of derivatives of $P$ and $Q$. We illustrate this with the help of a few examples. Consider, for instance, the term $\text{Im}  \text{Tr} [H_0(\partial_b   P) \mathcal Q^{(1)} (\partial_c  P)] $, which appears in \eqref{eq:dM_a}. We may use that 
\be
\mathcal Q^{(1)}  = -\mathcal P^{(1)} = -P\mathcal P^{(1)} Q-Q\mathcal P^{(1)} P, \label{app:P(1)}
\ee
which expresses the fact that $\mathcal P^{(1)}$ only has matrix elements between the occupied and unoccupied subspace. Substituting this into the term considered yields two terms, 
\be
\text{Im}  \text{Tr} [H_0(\partial_b   P) \mathcal Q^{(1)} (\partial_c  P)]  = 
-\text{Im}  \text{Tr} [H_0(\partial_b   P)P\mathcal P^{(1)} Q (\partial_c  P)]-\text{Im}  \text{Tr} [H_0(\partial_b   P) Q\mathcal P^{(1)} P(\partial_c  P)]
\ee
which can each be shown to vanish. In the first term we can rewrite $(\partial_b   P)P = -(\partial_b   Q)P = Q(\partial_b   P)$, where we have used $P+Q=\mathbb{1}$ and $QP=0$, and we can similarly rewrite $Q(\partial_c  P)=-(\partial_c  Q)P$. This then yields
\be
-\text{Im}  \text{Tr} [H_0(\partial_b   P)P\mathcal P^{(1)} Q (\partial_c  P)] =
\text{Im}  \text{Tr} [PH_0Q(\partial_b   P)\mathcal P^{(1)} (\partial_c  Q)] = 0
\ee
where in the last step we notice that $PH_0Q=0$ by definition. 

Similar manipulations lead to the conclusion that all remaining terms indeed vanish. For instance, as seen from Eq.~\eqref{app:int-parts}, integration by parts produces a term $
\text{Im}  \text{Tr} [H_0 \mathcal P^{(1)}(\partial_b   Q) (\partial_c  P)] $. Making the substitution $\mathcal P^{(1)} = P\mathcal P^{(1)} Q+Q\mathcal P^{(1)} P$ and successively making use of $Q(\partial_c  P) = -(\partial_c  Q)P$, as well as similar relations, shows that this term must vanish.

\subsection{Alternative derivation}

Next, we present an alternative but insightful derivation of the orbital magnetic polarizability, which closely follows the approach taken in Ref.~\onlinecite{Malashevich:2010p053032}. This derivation is more directly based on the Bloch eigenstates of the occupied bands, rather than the projectors constructed from them. Such derivation must therefore be prefaced with a more explicit discussion of gauge invariance and gauge transformations, so as to highlight the distinction between an arbitrary gauge choice and the so-called Hamiltonian gauge. 

A gauge transformation within the manifold of occupied bands is given by the unitary transformation
\be
|u_{n'} \rangle \rightarrow \sum_{n} U_\bk^{nn'} | u_n \rangle . \label{app:U-gauge}
\ee
The orbital magnetization is invariant under such a gauge transformation, and in order to make this gauge invariance manifest, it is necessary to assume an arbitrary gauge. Let $|   \bar u_{n} \rangle$ denote an arbitrary gauge choice. In an arbitrary gauge the matrix elements with the Hamiltonian are given by
\be
\mathcal H ^{nn'} = \langle  \bar u_n | \mathcal H | \bar u_{n'} \rangle. \label{app:H_nn'}
\ee
The Hamiltonian gauge is the special gauge for which $\mathcal H  | u_n \rangle  =  \mathcal E_n | u_n \rangle$ and in this (usual) gauge \eqref{app:H_nn'} reduces to $\mathcal H ^{nn'} =  \mathcal E_n \delta_{nn'}$. In what follows we work in an arbitrary gauge.

It is important to recall and emphasize that $\mathcal H$ in \eqref{app:H_nn'} is the full Hamiltonian (including the perturbation), such that the states $\{| \bar u_{n} \rangle\}$ have a dependence on the perturbation $\lambda$. In order to express the orbital magnetization in a manifestly gauge invariant way, we also introduce three other objects constructed from the eigenstates as \cite{Malashevich:2010p053032}
\be
 \mathcal  G^{nn'}_{ab} = \langle \nabla_a \bar u_n | \nabla_b \bar u_{n'} \rangle , \qquad \mathcal  H^{nn'}_{a}  = i \langle \bar u_n | \mathcal  H |\nabla_a \bar u_{n'} \rangle , \qquad
\mathcal   H^{nn'}_{ ab}=  \langle \nabla_a\bar  u_n | \mathcal  H |\nabla_b\bar u_{n'} \rangle. \label{app:quantities}
\ee
Here $ | \nabla_a \bar u_n \rangle $ denotes the covariant momentum derivative defined as
\be
 | \nabla_a \bar u_n \rangle  \equiv  \mathcal Q | \partial_a \bar u_n \rangle = (\mathbb{1} -\mathcal  P)| \partial_a \bar u_n \rangle, 
\ee
which is defined such that it transforms in a gauge covariant way. The first object given in \eqref{app:quantities} is known as the quantum geometric tensor. 
 
The quantities introduced in \eqref{app:H_nn'} and \eqref{app:quantities} can be regarded as matrices with matrix elements labeled by $n$ and $n'$, i.e., the occupied bands. The orbital magnetization can then be expressed as
\be
M_a = \frac{e}{2\hbar} \epsilon_{abc} \int [d\bk]~ \text{Im}\,  \text{tr} [\mathcal  H_{bc} + \mathcal  H \mathcal  G_{bc} ],
\ee
where $\text{tr} $ denotes the trace over the occupied subspace. Gauge invariance is then manifest, since the integrand is a trace over gauge covariant objects. 

The polarizability is then computed by taking the derivative with respect to $\lambda$, which yields
\be
 \frac{\partial M_a}{\partial \lambda}= \frac{e}{2\hbar} \epsilon_{abc} \int [d\bk] ~\text{Im}\,  \text{tr} [\partial_\lambda \mathcal H_{bc}    +\mathcal G_{bc}(\partial_\lambda  \mathcal H)+  \mathcal H (\partial_\lambda \mathcal G_{bc})]. \label{app:D_lambda-M}
\ee
To obtain an expression for the polarizability the three terms of the integrand must be evaluated. Taking the last two terms together, one finds
\begin{multline}
\epsilon_{abc} \int [d\bk] ~\text{Im}\,  \text{tr} [\mathcal G_{bc}(\partial_\lambda  \mathcal H)+  \mathcal H (\partial_\lambda \mathcal G_{bc})] \\
=\epsilon_{abc} \int [d\bk] \left(2 \text{Re}\,  \text{tr}[ \mathcal  A_\lambda \mathcal  H \mathcal  G_{bc} ]+2 \text{Re}~ \text{tr} [ \mathcal H \mathcal G_{b\lambda} \mathcal A_c ] 
+ \text{Im}~ \text{tr} [\mathcal  G_{bc}W] +2 \text{Im}\sum_{nn'} \mathcal H^{nn'} \langle  \partial_\lambda \partial_b \bar u_{n'} | \mathcal Q| \partial_c \bar u_{n} \rangle\right) \\
=  \epsilon_{abc} \int [d\bk] \bigg(  \text{Im}\, \text{tr} [ \mathcal G_{bc}W]  -2 \text{Im}\, \text{tr} [(\partial_b H_0) \mathcal G_{\lambda c}]   \bigg) , \label{app:reduce-1}
\end{multline}
where we have introduced the Berry connections
\be
\mathcal A^{nn'}_{\lambda} = i \langle \bar u_n | \partial_\lambda \bar u_{n'} \rangle,  \qquad \mathcal A^{nn'}_{a} = i \langle \bar u_n | \partial_a \bar u_{n'} \rangle,
\ee
and have made use of 
\be
\partial_\lambda \mathcal  H^{nn'}  = i [\mathcal  A_\lambda , \mathcal  H ]^{nn'} + W^{nn'}. \label{eq:D_lambda-H} 
\ee
To obtain the final result of Eq.~\eqref{app:reduce-1}, we have performed an integration by parts on the last term in the second line. Integration by parts produces a number of terms, which cancel most other terms except for the two terms of the final line~\cite{Malashevich:2010p053032}. 

Using similar manipulations, one finds that the first term in the integrand of \eqref{app:D_lambda-M} ultimately reduces to
\be
 \epsilon_{abc} \int [d\bk] ~\text{Im}\,  \text{tr} [\partial_\lambda  \mathcal H_{bc}]  =\epsilon_{abc} \int [d\bk] \Big(  \text{Im}\, \text{tr} [W_{bc}]  -2 \text{Im}\, \text{tr} [(\partial_b H_0)_{\lambda c}]   \Big) ,
\ee
where we have defined $W^{nn'}_{bc} = \langle \nabla_b\bar  u_n |W |\nabla_c\bar u_{n'} \rangle$ [in analogy with $\mathcal H^{nn'}_{bc}$ of \eqref{app:quantities}] and $(\partial_b H_0)^{nn'}_{\lambda c}=\langle \nabla_\lambda\bar  u_n | (\partial_b H_0) |\nabla_c\bar u_{n'} \rangle$. Here $\partial_b H_0$ should be read as the operator derivative of $H_0$ (and not the derivative of matrix elements of $H_0$ in the basis $|  \bar u_n \rangle$).  

To determine the polarizability from the derivative of $M_a$, we now seek to evaluate the derivative at $\lambda=0$. We group the four obtained terms into two contributions. The first contribution collects the two terms $  \text{Im}\, \text{tr} [ \mathcal G_{bc}W]$ and $\text{Im}\, \text{tr} [W_{bc}] $, and is straightforward to evaluate. We find
\begin{multline}
\epsilon_{abc} \int [d\bk] \Big(   \text{Im}\, \text{tr} [ \mathcal G_{bc}W]+\text{Im}\, \text{tr} [W_{bc}]   \Big)\Big|_{\lambda=0} \\
= \epsilon_{abc} \int [d\bk]   \text{Im} \Big(\sum_{nn'} \langle \partial_b u^0_{n} |Q| \partial_c u^0_{n'} \rangle \langle u^0_{n'} |W|u^0_n \rangle+\sum_n\langle \partial_b u^0_{n} |QWQ| \partial_c u^0_n \rangle   \Big). \label{app:geometric}
\end{multline}
To obtain the right hand side, we have simply evaluated $\mathcal G^{nn'}_{bc}$ and $W^{nn'}_{bc}$ in the unperturbed basis $|  u^0_n \rangle $ and adopted the Hamiltonian gauge. It is then possible to show that the right hand side of \eqref{app:geometric} exactly corresponds to the geometric term of Eq.~\eqref{eq:dM_a-final}.

Now consider the remaining terms, which involve derivatives of the Hamiltonian $H_0$. Writing them out explicitly and adopting the Hamiltonian gauge yields
\begin{multline}
-2\epsilon_{abc} \int [d\bk] \Big(\text{Im}\, \text{tr} [(\partial_b H_0) \mathcal G_{\lambda c}+\text{Im}\, \text{tr} [(\partial_b H_0)_{\lambda c}]    \Big)\Big|_{\lambda=0} \\
= -2 \epsilon_{abc} \int [d\bk]   \text{Im} \Big(\sum_{nn'} \langle \partial_b u_{n} |Q| \partial_\lambda u_{n'} \rangle \langle u_{n'} |(\partial_c H_0 )|u_n \rangle+\sum_n\langle \partial_b u_{n} |Q (\partial_c H_0 ) Q| \partial_\lambda u_n \rangle   \Big) \Big|_{\lambda=0} . \label{app:interband}
\end{multline} 
In the Hamiltonian gauge we can now apply standard perturbation theory (see Sec.~\ref{sec:problem}) to determine $| \partial_\lambda u_{n'} \rangle |_{\lambda=0}$, which is simply the first order correction to the eigenstates. After straightforward manipulations this contribution reduces to the interband term of Eq.~\eqref{eq:dM_a-final}.


\section{Analysis of interband term in Eqs.~\eqref{eq:dA_dBa} and \eqref{eq:dM_a-final} \label{app:interband}}

This Appendix presents additional analysis of the interband term appearing both in Eq.~\eqref{eq:dA_dBa} and in Eq.~\eqref{eq:dM_a-final}. The main goal is to demonstrate that the interband term vanishes when the spectrum of $H_0$ has the properties of ``degeneracy'' and ``reflection''.

We first show that the interband term can be expressed in terms of derivatives of the Hamiltonian only. To this end, we rewrite derivatives of the projectors as derivatives of the Hamiltonian. Note that 
\be
\partial_a  P = \nabla_a P = \sum_{n}( Q | \partial_a u^0_n \rangle \langle  u^0_n| + | u^0_n \rangle \langle  \partial_a u^0_n|Q)
\ee
where we recall that $\nabla_a $ denotes the covariant derivative, and $Q = \sum_m| u^0_m \rangle \langle u^0_m | $ is the projector onto the unoccupied bands. It is then straightforward to show that 
\be
\partial_a  P =\sum_{n,m}\frac{P_n (\partial_aH_0)P_m+P_m (\partial_aH_0)P_n }{\varepsilon_n-\varepsilon_m},
\ee
where (in keeping with notation used throughout this work) $n$ and $m$ denote sums over occupied and unoccupied bands, respectively. Substituting the projector derivative in the numerator of the interband term yields
\be
\text{Im}  \text{Tr} [ WP_n \{ \partial_bH_0  ,\partial_c  P\}P_m] = \sum_{n'}\frac{ \text{Im}  \text{Tr} [ WP_n (\partial_bH_0)P_{n'}( \partial_c  H_0)P_m] }{\varepsilon_{n'}-\varepsilon_m} +  \sum_{m'}\frac{ \text{Im}  \text{Tr} [ WP_n (\partial_cH_0)P_{m'}( \partial_b  H_0)P_m] }{\varepsilon_{n}-\varepsilon_{m'}}. \label{app:interband-rewrite}
\ee
One may then use the definition of velocity, i.e., $ v^{nn'}_a =  \langle  u^0_n|\partial_a H_0 | u^0_{n'} \rangle$ (similarly for other matrix elements), to write the interband term as
\be
\sum_{n,m}\frac{\text{Im}  \text{Tr} [ WP_n \{ \partial_bH_0  ,\partial_c  P\}P_m]}{\varepsilon_n-\varepsilon_m} 
= \sum_{nn'm}\frac{\text{Im}[ v^{nn'}_b v^{n'm}_c W^{mn} ] }{(\varepsilon_n-\varepsilon_m)(\varepsilon_{n'}-\varepsilon_m)}+\sum_{nmm'}\frac{\text{Im}[ v^{nm'}_c v^{m'm}_b W^{mn} ] }{(\varepsilon_n-\varepsilon_m)(\varepsilon_{n}-\varepsilon_{m'})}.
 \label{app:interband-velocity}
\ee

Next, we take the right hand side of Eq.~\eqref{app:interband-rewrite} and rewrite it entirely in terms of projectors and energies and their derivatives. Writing the Hamiltonian as
\be
H_0 = \sum_n \varepsilon_nP_n +  \sum_m \varepsilon_m P_m,
\ee
yields an expression for its derivative given by
\be
\partial_a H_0 = \sum_n (\partial_a\varepsilon_n)P_n+\sum_m (\partial_a\varepsilon_m)P_m+   \sum_n \varepsilon_n\partial_aP_n  +\sum_m\varepsilon_n\partial_aP_m. 
\ee
Using this form of the Hamiltonian derivative it is straightforward to show that
\begin{align}
P_m  (\partial_aH_0 )P_{m'}  &=P_m (\partial_a\varepsilon_{ m})\delta_{mm'}  -(\varepsilon_{ m}-\varepsilon_{m'}) P_m(\partial_a P_{m'}), \\
P_n ( \partial_a H_0 )P_{n'}  &= P_n (\partial_a\varepsilon_{ n})\delta_{nn'}  -(\varepsilon_{ n}-\varepsilon_{n'}) P_n(\partial_a P_{n'}), \\
P_n  (\partial_aH_0 ) P_m  &= -(\varepsilon_{ n}-\varepsilon_{m})P_n (\partial_a P_m), \label{eq:Pn-D-Pm} 
\end{align}
and these relations can  in turn be used to demonstrate that Eq.~\eqref{app:interband-rewrite} takes the form
\begin{multline}
\text{Im}  \text{Tr} [ WP_n \{ \partial_bH_0  ,\partial_c  P\}P_m] = - \partial_b(\varepsilon_{n}+\varepsilon_m)\text{Im}  \text{Tr} [ WP_n \partial_c  P_m]+ \sum_{n'}(\varepsilon_{n}-\varepsilon_{n'})\text{Im}  \text{Tr} [ WP_n(\partial_b P_{n'} )(\partial_c  P_m)] \\
+ \sum_{m'} (\varepsilon_{m'}-\varepsilon_m)\text{Im}  \text{Tr} [ WP_n(\partial_c P_{m'} )(\partial_b  P_m)]. \label{app:interband-final}
\end{multline}
Based on this expression we can now draw the following conclusions, depending on two properties which the spectrum of the unperturbed Hamiltonian $H_0$ may satisfy. The two properties of interest are: ({\it i}) (``degeneracy'') all occupied valence bands have the same energy and all unoccupied conduction bands have the same energy; ({\it ii}) (``reflection'') the sum of valence and conduction band energies, given by $\varepsilon_{n}+\varepsilon_m$, is independent of $\bk$. We consider two cases: the case when $H_0$ satisfies both degeneracy and reflection, and the case when $H_0$ satisfies only degeneracy but not reflection.

{\it Degeneracy and reflection.} When the spectrum satisfies both degeneracy and reflection the interband contribution vanishes. The vanishing of a contribution to polarizability was previously observed in the context of the orbital magnetoelectric polarizability~\cite{Essin:2010p205104}. Property ({\it i}) implies that both $\varepsilon_{n}-\varepsilon_{n'}$ and $\varepsilon_{m}-\varepsilon_{m'}$ vanish for all combinations of $n,n'$ and $m,m'$, since all valence bands have the same energy, as do the conduction bands. Property ({\it ii}) implies that the momentum derivative $\partial_b(\varepsilon_{n}+\varepsilon_m)$ vanishes. Therefore, the right hand side of \eqref{app:interband-final} vanishes.

{\it Only degeneracy.} When the spectrum of $H_0$ only satisfies degeneracy but not reflection, the interband does not vanish, but reduces to a form with an appealing interpretation. Degeneracy still implies that the last two terms in Eq.~\eqref{app:interband-final} vanish, but the absence of reflection means the first term in~\eqref{app:interband-final} may give a nonzero contribution. Since all valence bands have the same energy and all conduction bands have the same energy, we can write $\varepsilon_{n}+\varepsilon_m \equiv 2h_0$ for all $n$ and $m$, where $h_0$ is some function of momentum describing the uniform piece of the dispersion.  The interband term can then be written as
\be
\int[d\bk] \sum_{n,m} \frac{ \text{Im}  \text{Tr} [WP_n \{ \partial_b  H_0,\partial_c  P\}P_m    ] }{\varepsilon_n-\varepsilon_m} \rightarrow - \int[d\bk] \sum_{n,m}(\partial_b\varepsilon_0) \frac{2 \text{Im}  \text{Tr} [ WP_n \partial_c  P_m] }{\varepsilon_n-\varepsilon_m} = -\int[d\bk] (\partial_b\varepsilon_0) \Omega_{c\lambda} \big|_{\lambda=0},
\ee
which shows that the integrand term (under the assumption of degeneracy) is proportional to the generalized curvature $\Omega_{c\lambda} $. Integration by parts further relates the interband term to the Berry curvature polarizability, as discussed in Sec.~\ref{sec:Hall}.

\end{widetext}

\section{Proof of Eq.~\eqref{eq:Berry-sum}  \label{app:dOmega_ab}}

In this Appendix we provide details pertaining to the derivation of Eq.~\eqref{eq:Berry-sum} in Sec.~\ref{sec:Hall}. To establish Eq.~\eqref{eq:Berry-sum}, we simply take the appropriate derivatives of the curvatures and sum the results. For $\partial_\lambda \Omega_{ab}$ we find
\begin{multline}
\partial_\lambda \Omega_{ab} =  -2\text{Im}\text{Tr}[\partial_\lambda \mathcal P\partial_a \mathcal P \partial_b \mathcal P] -  2\text{Im}\text{Tr}[ \mathcal P \partial_\lambda\partial_a \mathcal P \partial_b \mathcal P] \\
 - 2\text{Im}\text{Tr}[  \mathcal P\partial_a \mathcal P \partial_\lambda\partial_b \mathcal P]  , \label{app:dOmega_ab}
\end{multline}
and cyclic permutation of the indices then yields
\begin{multline}
\partial_a \Omega_{b\lambda} =   -2\text{Im}\text{Tr}[\partial_a \mathcal P \partial_b \mathcal P\partial_\lambda \mathcal P] -  2\text{Im}\text{Tr}[ \mathcal P \partial_a\partial_b \mathcal P \partial_\lambda \mathcal P] \\
-2 \text{Im}\text{Tr}[  \mathcal P\partial_b \mathcal P \partial_a\partial_\lambda  \mathcal P] , \label{app:dOmega_b-lambda}
\end{multline}
as well as
\begin{multline}
\partial_b \Omega_{\lambda a} =   -2\text{Im}\text{Tr}[ \partial_b \mathcal P\partial_\lambda \mathcal P \partial_a \mathcal P] -  2\text{Im}\text{Tr}[ \mathcal P \partial_b\partial_\lambda \mathcal P \partial_a \mathcal P] \\
-2 \text{Im}\text{Tr}[  \mathcal P\partial_\lambda \mathcal P \partial_a\partial_b  \mathcal P]  . \label{app:dOmega-lambda-a}
\end{multline}
We then use that the trace of three projector derivatives vanishes identically, i.e., 
\be
\text{Tr}[\partial_\lambda \mathcal P\partial_a \mathcal P \partial_b \mathcal P] = 0,
\ee
which may be proven by inserting versions of $\partial \mathcal P = (\partial \mathcal P) \mathcal P+ \mathcal P\partial \mathcal P$ successively~\cite{Mitscherling:2025p085104}.

Taking the sum $\partial_\lambda \Omega_{ab}+\partial_a \Omega_{b\lambda} +\partial_b \Omega_{\lambda a} $, the remaining six terms can be paired to form three pairs, which each can be shown to vanish. For instance, one has
\be
\text{Im}\text{Tr}[ \mathcal P \partial_\lambda\partial_a \mathcal P \partial_b \mathcal P] + \text{Im}\text{Tr}[  \mathcal P\partial_b \mathcal P \partial_a\partial_\lambda  \mathcal P] =0
\ee
which follows from the identity $\text{Im}~z  + \text{Im}~z^* =0$ and the cyclic property of the trace. We thus arrive at Eq.~\eqref{eq:Berry-sum} of the main text. To further arrive at Eq.~\eqref{eq:dOmega_ab}, we use the perturbative result of Sec.~\ref{ssec:Berry-P}:
\begin{align}
\Omega_{a\lambda} &=  -2 \text{Im}\text{Tr}[   P\partial_a   P  \mathcal P^{(1)}] , \\
&= \sum_{n,m}\frac{2\text{Im}\text{Tr}[   P_n(\partial_a   P_m)W  ] }{\varepsilon_{n}-\varepsilon_m}.
\end{align}

\end{document}